%Paper: hep-th/9402048
%From: okumura_yoshitaka <hiromu@isc.chubu.ac.jp>
%Date: Wed, 9 Feb 1994 14:23:16 +0900

\catcode`@=11 % This allows us to modify PLAIN macros.
%\font\twelverm=cmr10 scaled\magstep1
%\font\ninerm=cmr9	     \font\sixrm=cmr6
%AOKI

\font\fourteenrm=cmr10 scaled\magstep2
\font\twelverm=cmr12

\font\ninerm=cmr9

\font\sixrm=cmr6

\font\fourteenbf=cmbx10 scaled\magstep2
\font\twelvebf=cmbx10 scaled\magstep1
\font\ninebf=cmbx9	      \font\sixbf=cmbx6
\font\seventeeni=cmmi10 scaled\magstep3	    \skewchar\seventeeni='177
\font\fourteeni=cmmi10 scaled\magstep2	    \skewchar\fourteeni='177
%IKEMORI
%\font\twelvei=cmmi10 scaled\magstep1        \skewchar\twelvei='177
\font\twelvei=cmmi12                        \skewchar\twelvei='177
\font\ninei=cmmi9			    \skewchar\ninei='177
\font\sixi=cmmi6			    \skewchar\sixi='177
\font\seventeensy=cmsy10 scaled\magstep3    \skewchar\seventeensy='60
\font\fourteensy=cmsy10 scaled\magstep2	    \skewchar\fourteensy='60
\font\twelvesy=cmsy10 scaled\magstep1	    \skewchar\twelvesy='60
\font\ninesy=cmsy9			    \skewchar\ninesy='60
\font\sixsy=cmsy6			    \skewchar\sixsy='60

\font\fourteenex=cmex10 scaled\magstep2
\font\twelveex=cmex10 scaled\magstep1

\font\fourteensl=cmsl10 scaled\magstep2
\font\twelvesl=cmsl10 scaled\magstep1

\font\ninesl=cmsl9

\font\fourteenit=cmti10 scaled\magstep2
\font\twelveit=cmti10 scaled\magstep1
\font\twelvett=cmtt10 scaled\magstep1

%
%AOKI

\font\fourteencp=cmcsc10 scaled\magstep2
\font\twelvecp=cmcsc10 scaled\magstep1
\font\tencp=cmcsc10
\newfam\cpfam
%
	% quick fix for a missing font
%
\newcount\f@ntkey	     \f@ntkey=0
\def\samef@nt{\relax \ifcase\f@ntkey \rm \or\oldstyle \or\or
	 \or\it \or\sl \or\bf \or\tt \or\caps \fi }
%
%IKEMORI % fonts for jtex
\jfont\tenmin=min10
\jfont\twelvemin=min10 scaled\magstep1
\jfont\fourteenmin=min10 scaled\magstep2
\jfont\seventeenmin=min10 scaled\magstep3
\jfont\tengt=goth10
\jfont\twelvegt=goth10 scaled\magstep1
\jfont\fourteengt=goth10 scaled\magstep2
\jfont\seventeengt=goth10 scaled\magstep3
\def\fourteenpoint{\relax
    \textfont0=\fourteenrm
    \scriptfont0=\tenrm
    \scriptscriptfont0=\sevenrm
     \def\rm{\fam0 \fourteenrm \f@ntkey=0 }\relax
    \textfont1=\fourteeni	    \scriptfont1=\teni
    \scriptscriptfont1=\seveni
     \def\oldstyle{\fam1 \fourteeni\f@ntkey=1 }\relax
    \textfont2=\fourteensy	    \scriptfont2=\tensy
    \scriptscriptfont2=\sevensy
    \textfont3=\fourteenex     \scriptfont3=\fourteenex
    \scriptscriptfont3=\fourteenex
    \def\it{\fam\itfam \fourteenit\f@ntkey=4 }\textfont\itfam=\fourteenit
    \def\sl{\fam\slfam \fourteensl\f@ntkey=5 }\textfont\slfam=\fourteensl
    \scriptfont\slfam=\tensl
    \def\bf{\fam\bffam \fourteenbf\f@ntkey=6 }\textfont\bffam=\fourteenbf
    \scriptfont\bffam=\tenbf	 \scriptscriptfont\bffam=\sevenbf
    \def\tt{\fam\ttfam \twelvett \f@ntkey=7 }\textfont\ttfam=\twelvett
    \h@big=11.9\p@{} \h@Big=16.1\p@{} \h@bigg=20.3\p@{} \h@Bigg=24.5\p@{}
    \def\caps{\fam\cpfam \twelvecp \f@ntkey=8 }\textfont\cpfam=\twelvecp
    \setbox\strutbox=\hbox{\vrule height 12pt depth 5pt width\z@}
    \samef@nt}
\def\twelvepoint{\relax
    \textfont0=\twelverm
    \scriptfont0=\ninerm
    \scriptscriptfont0=\sixrm
     \def\rm{\fam0 \twelverm \f@ntkey=0 }\relax
    \textfont1=\twelvei            \scriptfont1=\ninei
    \scriptscriptfont1=\sixi
     \def\oldstyle{\fam1 \twelvei\f@ntkey=1 }\relax
    \textfont2=\twelvesy	  \scriptfont2=\ninesy
    \scriptscriptfont2=\sixsy
    \textfont3=\twelveex	  \scriptfont3=\twelveex
    \scriptscriptfont3=\twelveex
    \def\it{\fam\itfam \twelveit \f@ntkey=4 }\textfont\itfam=\twelveit
    \def\sl{\fam\slfam \twelvesl \f@ntkey=5 }\textfont\slfam=\twelvesl
    \scriptfont\slfam=\ninesl
    \def\bf{\fam\bffam \twelvebf \f@ntkey=6 }\textfont\bffam=\twelvebf
    \scriptfont\bffam=\ninebf	  \scriptscriptfont\bffam=\sixbf
    \def\tt{\fam\ttfam \twelvett \f@ntkey=7 }\textfont\ttfam=\twelvett
    \h@big=10.2\p@{}
    \h@Big=13.8\p@{}
    \h@bigg=17.4\p@{}
    \h@Bigg=21.0\p@{}
    \def\caps{\fam\cpfam \twelvecp \f@ntkey=8 }\textfont\cpfam=\twelvecp
    \setbox\strutbox=\hbox{\vrule height 10pt depth 4pt width\z@}
    \samef@nt}
\def\tenpoint{\relax
    \textfont0=\tenrm
    \scriptfont0=\sevenrm
    \scriptscriptfont0=\fiverm
    \def\rm{\fam0 \tenrm \f@ntkey=0 }\relax
    \textfont1=\teni	       \scriptfont1=\seveni
    \scriptscriptfont1=\fivei
    \def\oldstyle{\fam1 \teni \f@ntkey=1 }\relax
    \textfont2=\tensy	       \scriptfont2=\sevensy
    \scriptscriptfont2=\fivesy
    \textfont3=\tenex	       \scriptfont3=\tenex
    \scriptscriptfont3=\tenex
    \def\it{\fam\itfam \tenit \f@ntkey=4 }\textfont\itfam=\tenit
    \def\sl{\fam\slfam \tensl \f@ntkey=5 }\textfont\slfam=\tensl
    \def\bf{\fam\bffam \tenbf \f@ntkey=6 }\textfont\bffam=\tenbf
    \scriptfont\bffam=\sevenbf	   \scriptscriptfont\bffam=\fivebf
    \def\tt{\fam\ttfam \tentt \f@ntkey=7 }\textfont\ttfam=\tentt
    \def\caps{\fam\cpfam \tencp \f@ntkey=8 }\textfont\cpfam=\tencp
    \setbox\strutbox=\hbox{\vrule height 8.5pt depth 3.5pt width\z@}
    \samef@nt}
%
%%%%%%%%%%%%%%%%%%%%%%%%%%%%%%%%%%%%%%%%%%%%%%%%%%%%%%%%%%%%%%%%%%%%%%%%
%
%   Next redifine \big \Big \bigg and \Bigg to work with all fonts
%
%%%%%%%%%%%%%%%%%%%%%%%%%%%%%%%%%%%%%%%%%%%%%%%%%%%%%%%%%%%%%%%%%%%%%%%%
%
\newdimen\h@big  \h@big=8.5\p@
\newdimen\h@Big  \h@Big=11.5\p@
\newdimen\h@bigg  \h@bigg=14.5\p@
\newdimen\h@Bigg  \h@Bigg=17.5\p@
\def\big#1{{\hbox{$\left#1\vbox to\h@big{}\right.\n@space$}}}
\def\Big#1{{\hbox{$\left#1\vbox to\h@Big{}\right.\n@space$}}}
\def\bigg#1{{\hbox{$\left#1\vbox to\h@bigg{}\right.\n@space$}}}
\def\Bigg#1{{\hbox{$\left#1\vbox to\h@Bigg{}\right.\n@space$}}}
%
%%%%%%%%%%%%%%%%%%%%%%%%%%%%%%%%%%%%%%%%%%%%%%%%%%%%%%%%%%%%%%%%%%%%%%%%
%
%   Next, I define basic spacing parameters.
%
%\normalbaselineskip = 20pt plus 0.2pt minus 0.1pt
\normalbaselineskip = 18pt plus 0.2pt minus 0.1pt
\normallineskip = 1.5pt plus 0.1pt minus 0.1pt
\normallineskiplimit = 1.5pt
\newskip\normaldisplayskip
\normaldisplayskip = 20pt plus 5pt minus 10pt
\newskip\normaldispshortskip
\normaldispshortskip = 6pt plus 5pt
\newskip\normalparskip
\normalparskip = 6pt plus 2pt minus 1pt
\newskip\skipregister
\skipregister = 5pt plus 2pt minus 1.5pt
\newif\ifsingl@	   \newif\ifdoubl@
\newif\iftwelv@	   \twelv@true
\def\singlespace{\singl@true\doubl@false\spaces@t}
\def\doublespace{\singl@false\doubl@true\spaces@t}
\def\normalspace{\singl@false\doubl@false\spaces@t}
\def\Tenpoint{\tenpoint\twelv@false\spaces@t}
\def\Twelvepoint{\twelvepoint\twelv@true\spaces@t}
\def\spaces@t{\relax%
 \iftwelv@ \ifsingl@\subspaces@t3:4;\else\subspaces@t1:1;\fi%
 \else \ifsingl@\subspaces@t3:5;\else\subspaces@t4:5;\fi \fi%
 \ifdoubl@ \multiply\baselineskip by 5%
 \divide\baselineskip by 4 \fi \unskip}
\def\subspaces@t#1:#2;{%
      \baselineskip = \normalbaselineskip%
      \multiply\baselineskip by #1 \divide\baselineskip by #2%
      \lineskip = \normallineskip%
      \multiply\lineskip by #1 \divide\lineskip by #2%
      \lineskiplimit = \normallineskiplimit%
      \multiply\lineskiplimit by #1 \divide\lineskiplimit by #2%
      \parskip = \normalparskip%
      \multiply\parskip by #1 \divide\parskip by #2%
      \abovedisplayskip = \normaldisplayskip%
      \multiply\abovedisplayskip by #1 \divide\abovedisplayskip by #2%
      \belowdisplayskip = \abovedisplayskip%
      \abovedisplayshortskip = \normaldispshortskip%
      \multiply\abovedisplayshortskip by #1%
	\divide\abovedisplayshortskip by #2%
      \belowdisplayshortskip = \abovedisplayshortskip%
      \advance\belowdisplayshortskip by \belowdisplayskip%
      \divide\belowdisplayshortskip by 2%
      \smallskipamount = \skipregister%
      \multiply\smallskipamount by #1 \divide\smallskipamount by #2%
      \medskipamount = \smallskipamount \multiply\medskipamount by 2%
      \bigskipamount = \smallskipamount \multiply\bigskipamount by 4 }
\def\normalbaselines{ \baselineskip=\normalbaselineskip%
   \lineskip=\normallineskip \lineskiplimit=\normallineskip%
   \iftwelv@\else \multiply\baselineskip by 4 \divide\baselineskip by 5%
     \multiply\lineskiplimit by 4 \divide\lineskiplimit by 5%
     \multiply\lineskip by 4 \divide\lineskip by 5 \fi }
\Twelvepoint  % That's the default
\interlinepenalty=50
\interfootnotelinepenalty=5000
\predisplaypenalty=9000
\postdisplaypenalty=500
\hfuzz=1pt
\vfuzz=0.2pt
%
%%%%%%%%%%%%%%%%%%%%%%%%%%%%%%%%%%%%%%%%%%%%%%%%%%%%%%%%%%%%%%%%%%%%%%%%
%
%   Next, I define output routines, footnotes & related stuff.
%
\def\pagecontents{%
   \ifvoid\topins\else\unvbox\topins\vskip\skip\topins\fi
   \dimen@ = \dp255 \unvbox255
   \ifvoid\footins\else\vskip\skip\footins\footrule\unvbox\footins\fi
   \ifr@ggedbottom \kern-\dimen@ \vfil \fi }
\def\makeheadline{\vbox to 0pt{ \skip@=\topskip
      \advance\skip@ by -12pt \advance\skip@ by -2\normalbaselineskip
      \vskip\skip@ \line{\vbox to 12pt{}\the\headline} \vss
      }\nointerlineskip}
\def\makefootline{\baselineskip = 1.5\normalbaselineskip
		 \line{\the\footline}}
\newif\iffrontpage
\newif\ifletterstyle
\newif\ifp@genum
\def\nopagenumbers{\p@genumfalse}
\def\pagenumbers{\p@genumtrue}
\pagenumbers
\newtoks\paperheadline
\newtoks\letterheadline
\newtoks\letterfrontheadline
\newtoks\lettermainheadline
\newtoks\paperfootline
\newtoks\letterfootline
\newtoks\date
\footline={\ifletterstyle\the\letterfootline\else\the\paperfootline\fi}
%\paperfootline={\hss\iffrontpage\else\ifp@genum\tenrm\folio\hss\fi\fi}
%AOKI
\paperfootline={\hss\iffrontpage\else\ifp@genum\tenrm
    -- \folio\ --\hss\fi\fi}
\letterfootline={\hfil}
\headline={\ifletterstyle\the\letterheadline\else\the\paperheadline\fi}
\paperheadline={\hfil}
\letterheadline{\iffrontpage\the\letterfrontheadline
     \else\the\lettermainheadline\fi}
\lettermainheadline={\rm\ifp@genum page \ \folio\fi\hfil\the\date}
\def\monthname{\relax\ifcase\month 0/\or January\or February\or
   March\or April\or May\or June\or July\or August\or September\or
   October\or November\or December\else\number\month/\fi}
\date={\monthname\ \number\day, \number\year}
\countdef\pagenumber=1  \pagenumber=1
\def\advancepageno{\global\advance\pageno by 1
   \ifnum\pagenumber<0 \global\advance\pagenumber by -1
    \else\global\advance\pagenumber by 1 \fi \global\frontpagefalse }
\def\folio{\ifnum\pagenumber<0 \romannumeral-\pagenumber
	   \else \number\pagenumber \fi }
\def\footrule{\dimen@=\prevdepth\nointerlineskip
   \vbox to 0pt{\vskip -0.25\baselineskip \hrule width 0.35\hsize \vss}
   \prevdepth=\dimen@ }
\newtoks\foottokens
\foottokens={\Tenpoint\singlespace}
\newdimen\footindent
\footindent=24pt
\def\vfootnote#1{\insert\footins\bgroup  \the\foottokens
   \interlinepenalty=\interfootnotelinepenalty \floatingpenalty=20000
   \splittopskip=\ht\strutbox \boxmaxdepth=\dp\strutbox
   \leftskip=\footindent \rightskip=\z@skip
   \parindent=0.5\footindent \parfillskip=0pt plus 1fil
   \spaceskip=\z@skip \xspaceskip=\z@skip
   \Textindent{$ #1 $}\footstrut\futurelet\next\fo@t}
\def\Textindent#1{\noindent\llap{#1\enspace}\ignorespaces}
\def\footnote#1{\attach{#1}\vfootnote{#1}}

\let\footsymbol=\star
\newcount\lastf@@t	     \lastf@@t=-1
\newcount\footsymbolcount    \footsymbolcount=0
\newif\ifPhysRev
\def\footsymbolgen{\relax \ifPhysRev \iffrontpage \NPsymbolgen\else
      \PRsymbolgen\fi \else \NPsymbolgen\fi
   \global\lastf@@t=\pageno \footsymbol }
\def\NPsymbolgen{\ifnum\footsymbolcount<0 \global\footsymbolcount=0\fi
   {\iffrontpage \else \advance\lastf@@t by 1 \fi
    \ifnum\lastf@@t<\pageno \global\footsymbolcount=0
     \else \global\advance\footsymbolcount by 1 \fi }
   \ifcase\footsymbolcount \fd@f\star\or \fd@f\dagger\or \fd@f\ast\or
    \fd@f\ddagger\or \fd@f\natural\or \fd@f\diamond\or \fd@f\bullet\or
    \fd@f\nabla\else \fd@f\dagger\global\footsymbolcount=0 \fi }
\def\fd@f#1{\xdef\footsymbol{#1}}
\def\PRsymbolgen{\ifnum\footsymbolcount>0 \global\footsymbolcount=0\fi
      \global\advance\footsymbolcount by -1
      \xdef\footsymbol{\sharp\number-\footsymbolcount} }
\def\space@ver#1{\let\@sf=\empty \ifmmode #1\else \ifhmode
   \edef\@sf{\spacefactor=\the\spacefactor}\unskip${}#1$\relax\fi\fi}
\def\attach#1{\space@ver{\strut^{\mkern 2mu #1} }\@sf\ }
%
%%%%%%%%%%%%%%%%%%%%%%%%%%%%%%%%%%%%%%%%%%%%%%%%%%%%%%%%%%%%%%%%%%%%%%%%
%
%   Here come chapter, section, subsection & appendix macros.
%
\newcount\chapternumber	     \chapternumber=0
\newcount\sectionnumber	     \sectionnumber=0
\newcount\equanumber	     \equanumber=0
\let\chapterlabel=0
\newtoks\chapterstyle	     \chapterstyle={\Number}
\newskip\chapterskip	     \chapterskip=\bigskipamount
\newskip\sectionskip	     \sectionskip=\medskipamount
\newskip\headskip	     \headskip=8pt plus 3pt minus 3pt
\newdimen\chapterminspace    \chapterminspace=15pc
\newdimen\sectionminspace    \sectionminspace=10pc
\newdimen\referenceminspace  \referenceminspace=25pc
\def\chapterreset{\global\advance\chapternumber by 1
   \ifnum\equanumber<0 \else\global\equanumber=0\fi
   \sectionnumber=0 \makel@bel}
\def\makel@bel{\xdef\chapterlabel{%
\the\chapterstyle{\the\chapternumber}.}}
\def\sectionlabel{\number\sectionnumber \quad }
\def\alphabetic#1{\count255='140 \advance\count255 by #1\char\count255}
\def\Alphabetic#1{\count255='100 \advance\count255 by #1\char\count255}
\def\Roman#1{\uppercase\expandafter{\romannumeral #1}}
\def\roman#1{\romannumeral #1}
\def\Number#1{\number #1}
\def\unnumberedchapters{\let\makel@bel=\relax \let\chapterlabel=\relax
\let\sectionlabel=\relax \equanumber=-1 }
%
%AOKI
\def\titlestyle#1{\par\begingroup \interlinepenalty=9999
     \leftskip=0.03\hsize plus 0.20\hsize minus 0.03\hsize
     \rightskip=\leftskip \parfillskip=0pt
     \hyphenpenalty=9000 \exhyphenpenalty=9000
     \tolerance=9999 \pretolerance=9000
     \spaceskip=0.333em \xspaceskip=0.5em
     \iftwelv@\fourteenpoint\fourteenbf
     \else\twelvepoint\twelvebf\fi
     \noindent  #1\par\endgroup }
\def\spacecheck#1{\dimen@=\pagegoal\advance\dimen@ by -\pagetotal
   \ifdim\dimen@<#1 \ifdim\dimen@>0pt \vfil\break \fi\fi}
\def\chapter#1{\par \penalty-300 \vskip\chapterskip
   \spacecheck\chapterminspace
   \chapterreset \titlestyle{\chapterlabel \ \fourteengt #1}
   \nobreak\vskip\headskip \penalty 30000
   \wlog{\string\chapter\ \chapterlabel } }
%AOKI

%
\def\section#1{\par \ifnum\the\lastpenalty=30000\else
   \penalty-200\vskip\sectionskip \spacecheck\sectionminspace\fi
   \wlog{\string\section\ \chapterlabel\the\sectionnumber}
   \global\advance\sectionnumber by 1  \noindent
   {\caps\enspace\chapterlabel\sectionlabel \twelvegt #1}\par
   \nobreak\vskip\headskip \penalty 30000 }
%AOKI
\def\ssection#1{\par \ifnum\the\lastpenalty=30000\else
   \penalty-200\vskip\sectionskip \spacecheck\sectionminspace\fi
   \wlog{\string\section\ \chapterlabel\the\sectionnumber}
   \global\advance\sectionnumber by 1  \noindent
   {\S \caps\thinspace\chapterlabel\sectionlabel #1}\par
   \nobreak\vskip\headskip \penalty 30000 }
\def\subsection#1{\par
   \ifnum\the\lastpenalty=30000\else \penalty-100\smallskip \fi
   \noindent\undertext{\twelvegt #1}\enspace \vadjust{\penalty5000}}

\def\undertext#1{\vtop{\hbox{#1}\kern 1pt \hrule}}
\def\APPENDIX#1#2{\par\penalty-300\vskip\chapterskip
   \spacecheck\chapterminspace \chapterreset \xdef\chapterlabel {#1}
   \titlestyle{APPENDIX #2} \nobreak\vskip\headskip \penalty 30000
   \wlog{\string\Appendix\ \chapterlabel } }
\def\Appendix#1{\APPENDIX{#1}{#1}}
\def\appendix{\APPENDIX{A}{}}
%
%%%%%%%%%%%%%%%%%%%%%%%%%%%%%%%%%%%%%%%%%%%%%%%%%%%%%%%%%%%%%%%%%%%%%%%%
%
%   Here come macros for equation numbering.
%
\def\eqname#1{\relax \ifnum\equanumber<0
     \xdef#1{{\rm(\number-\equanumber)}}\global\advance\equanumber by -1
    \else \global\advance\equanumber by 1
      \xdef#1{{\rm(\chapterlabel \number\equanumber)}} \fi}
\def\eqinsert#1{\noalign{\dimen@=\prevdepth \nointerlineskip
   \setbox0=\hbox to\displaywidth{\hfil #1}
   \vbox to 0pt{\vss\hbox{$\!\box0\!$}\kern-0.5\baselineskip}
   \prevdepth=\dimen@}}
%

%

%

%
%%%%%%%%%%%%%%%%%%%%%%%%%%%%%%%%%%%%%%%%%%%%%%%%%%%%%%%%%%%%%%%%%%%%%%%%
%   Here come items and lists
%
%\def\GENITEM#1;#2{\par \hangafter=0 \hangindent=#1
%    \Textindent{$ #2 $}\ignorespaces}
%AOKI
\def\GENITEM#1;#2{\par \hangafter=0 \hangindent=#1
    \Textindent{#2}\ignorespaces}
\outer\def\newitem#1=#2;{\gdef#1{\GENITEM #2;}}
\newdimen\itemsize		  \itemsize=30pt
\newitem\item=1\itemsize;
\newitem\sitem=1.75\itemsize;	  
\newitem\ssitem=2.5\itemsize;	  
\outer\def\newlist#1=#2&#3&#4;{\toks0={#2}\toks1={#3}%
   \count255=\escapechar \escapechar=-1
   \alloc@0\list\countdef\insc@unt\listcount	 \listcount=0
   \edef#1{\par
      \countdef\listcount=\the\allocationnumber
      \advance\listcount by 1
      \hangafter=0 \hangindent=#4
      \Textindent{\the\toks0{\listcount}\the\toks1}}
   \expandafter\expandafter\expandafter
    \edef\c@t#1{begin}{\par
      \countdef\listcount=\the\allocationnumber \listcount=1
      \hangafter=0 \hangindent=#4
      \Textindent{\the\toks0{\listcount}\the\toks1}}
   \expandafter\expandafter\expandafter
    \edef\c@t#1{con}{\par \hangafter=0 \hangindent=#4 \noindent}
   \escapechar=\count255}
\def\c@t#1#2{\csname\string#1#2\endcsname}
\newlist\point=\Number&.&1.0\itemsize;
\newlist\subpoint=(\alphabetic&)&1.75\itemsize;
\newlist\subsubpoint=(\roman&)&2.5\itemsize;
%

%
%%%%%%%%%%%%%%%%%%%%%%%%%%%%%%%%%%%%%%%%%%%%%%%%%%%%%%%%%%%%%%%%%%%%%%%%
%
%   Here come macros for references, figures & tables.
%
\newcount\referencecount     \referencecount=0
\newif\ifreferenceopen	     \newwrite\referencewrite
\newtoks\rw@toks
\def\NPrefmark#1{\attach{\scriptscriptstyle [ #1 ] }}
\let\PRrefmark=\attach
\def\refmark#1{\relax\ifPhysRev\PRrefmark{#1}\else\NPrefmark{#1}\fi}
\def\refend{\refmark{\number\referencecount}}
\newcount\lastrefsbegincount \lastrefsbegincount=0
\def\refsend{\refmark{\count255=\referencecount
   \advance\count255 by-\lastrefsbegincount
   \ifcase\count255 \number\referencecount
   \or \number\lastrefsbegincount,\number\referencecount
   \else \number\lastrefsbegincount-\number\referencecount \fi}}
\def\refch@ck{\chardef\rw@write=\referencewrite
   \ifreferenceopen \else \referenceopentrue
   \immediate\openout\referencewrite=reference.aux \fi}
%
% In \obeyendofline, we say `\let^^M=\relax
{\catcode`\^^M=\active % these lines must end with %
  \gdef\obeyendofline{\catcode`\^^M\active \let^^M\ }}%
%
% In \ignoreendofline, we say `\let^^M=\relax
{\catcode`\^^M=\active % these lines must end with %
  \gdef\ignoreendofline{\catcode`\^^M=5}}
{\obeyendofline\gdef\rw@start#1{\def\t@st{#1} \ifx\t@st\blankend%
\endgroup \@sf \relax \else \ifx\t@st\bl@nkend \endgroup \@sf \relax%
\else \rw@begin#1
\backtotext
\fi \fi } }
{\obeyendofline\gdef\rw@begin#1
{\def\n@xt{#1}\rw@toks={#1}\relax%
\rw@next}}
\def\blankend{}
{\obeylines\gdef\bl@nkend{
}}
\newif\iffirstrefline  \firstreflinetrue
\def\rwr@teswitch{\ifx\n@xt\blankend \let\n@xt=\rw@begin %
 \else\iffirstrefline \global\firstreflinefalse%
\immediate\write\rw@write{\noexpand\obeyendofline \the\rw@toks}%
\let\n@xt=\rw@begin%
      \else\ifx\n@xt\rw@@d \def\n@xt{\immediate\write\rw@write{%
	\noexpand\ignoreendofline}\endgroup \@sf}%
	     \else \immediate\write\rw@write{\the\rw@toks}%
	     \let\n@xt=\rw@begin\fi\fi \fi}
\def\rw@next{\rwr@teswitch\n@xt}
\def\rw@@d{\backtotext} \let\rw@end=\relax
\let\backtotext=\relax

\newdimen\refindent	\refindent=30pt
\def\refitem#1{\par \hangafter=0 \hangindent=\refindent \Textindent{#1}}
\def\REFNUM#1{\space@ver{}\refch@ck \firstreflinetrue%
 \global\advance\referencecount by 1 \xdef#1{\the\referencecount}}
\def\refnum#1{\space@ver{}\refch@ck \firstreflinetrue%
 \global\advance\referencecount by 1 \xdef#1{\the\referencecount}\refend}

\def\REF#1{\REFNUM#1%
 \immediate\write\referencewrite{%
 \noexpand\refitem{#1.}}%
\begingroup\obeyendofline\rw@start}
\def\ref{\refnum\?%
 \immediate\write\referencewrite{\noexpand\refitem{\?.}}%
\begingroup\obeyendofline\rw@start}
\def\Ref#1{\refnum#1%
 \immediate\write\referencewrite{\noexpand\refitem{#1.}}%
\begingroup\obeyendofline\rw@start}
\def\REFS#1{\REFNUM#1\global\lastrefsbegincount=\referencecount
\immediate\write\referencewrite{\noexpand\refitem{#1.}}%
\begingroup\obeyendofline\rw@start}
\newcount\figurecount	  \figurecount=0
\newif\iffigureopen	  \newwrite\figurewrite
\def\figch@ck{\chardef\rw@write=\figurewrite \iffigureopen\else
   \immediate\openout\figurewrite=figures.aux
   \figureopentrue\fi}
\def\FIGNUM#1{\space@ver{}\figch@ck \firstreflinetrue%
 \global\advance\figurecount by 1 \xdef#1{\the\figurecount}}
\def\FIG#1{\FIGNUM#1
   \immediate\write\figurewrite{\noexpand\refitem{#1.}}%
   \begingroup\obeyendofline\rw@start}
\def\fig{\FIGNUM\? fig.~\?%
\immediate\write\figurewrite{\noexpand\refitem{\?.}}%
\begingroup\obeyendofline\rw@start}
\def\figure{\FIGNUM\? figure~\?
   \immediate\write\figurewrite{\noexpand\refitem{\?.}}%
   \begingroup\obeyendofline\rw@start}
\def\Fig{\FIGNUM\? Fig.~\?%
\immediate\write\figurewrite{\noexpand\refitem{\?.}}%
\begingroup\obeyendofline\rw@start}
\def\Figure{\FIGNUM\? Figure~\?%
\immediate\write\figurewrite{\noexpand\refitem{\?.}}%
\begingroup\obeyendofline\rw@start}
\newcount\tablecount	 \tablecount=0
\newif\iftableopen	 \newwrite\tablewrite
\def\tabch@ck{\chardef\rw@write=\tablewrite \iftableopen\else
   \immediate\openout\tablewrite=tables.aux
   \tableopentrue\fi}
\def\TABNUM#1{\space@ver{}\tabch@ck \firstreflinetrue%
 \global\advance\tablecount by 1 \xdef#1{\the\tablecount}}
\def\TABLE#1{\TABNUM#1
   \immediate\write\tablewrite{\noexpand\refitem{#1.}}%
   \begingroup\obeyendofline\rw@start}
\def\Table{\TABNUM\? Table~\?%
\immediate\write\tablewrite{\noexpand\refitem{\?.}}%
\begingroup\obeyendofline\rw@start}
%

%
%%%%%%%%%%%%%%%%%%%%%%%%%%%%%%%%%%%%%%%%%%%%%%%%%%%%%%%%%%%%%%%%%%%%%%%%
%
%   Here come macros for memos & letters.
%
\def\masterreset{\global\pagenumber=1 \global\chapternumber=0
   \global\equanumber=0 \global\sectionnumber=0
   \global\referencecount=0 \global\figurecount=0 \global\tablecount=0 }
\def\FRONTPAGE{\ifvoid255\else\vfill\penalty-2000\fi
      \masterreset\global\frontpagetrue
      \global\lastf@@t=0 \global\footsymbolcount=0}

\def\paperstyle{\letterstylefalse\normalspace\papersize}
\def\letterstyle{\letterstyletrue\singlespace\lettersize}
%
%AOKI
\def\papersize{\hsize=35.2pc\vsize=52.7pc\hoffset=0.5pc\voffset=0.8pc
	       \skip\footins=\bigskipamount}
\def\lettersize{\hsize=35.2pc\vsize=50.0pc\hoffset=0.5pc\voffset=2.5pc
   \skip\footins=\smallskipamount \multiply\skip\footins by 3 }
\paperstyle   %  This is the default
%
% % % % % % % % % % % % % % % % % % % % % % % % % % % % % % % % % % % %
%
% 4/5/1989 IKEMORI
\def\MEMO{\letterstyle\FRONTPAGE \letterfrontheadline={\hfil}
    \line{\quad\fourteenrm NU MEMORANDUM\hfil\twelverm\the\date\quad}
    \medskip \memod@f}

\def\memit@m#1{\smallskip \hangafter=0 \hangindent=1in
      \Textindent{\caps #1}}
\def\memod@f{\xdef\to{\memit@m{To:}}\xdef\from{\memit@m{From:}}%
     \xdef\topic{\memit@m{Topic:}}\xdef\subject{\memit@m{Subject:}}%
     \xdef\rule{\bigskip\hrule height 1pt\bigskip}}
\memod@f
%
%AOKI 6/25/1986 The two lines above were modified as below by H.Mawatari

%IKEMORI 5/4/1989

%

%

%
%AOKI

%
\newskip\lettertopfil
\lettertopfil = -40pt plus 10pt minus 10pt
\newskip\letterbottomfil
\letterbottomfil = 0pt plus 2.3in minus 0pt
\newskip\spskip \setbox0\hbox{\ } \spskip=-1\wd0
\def\addressee#1{\medskip \rightline{\the\date\hskip 30pt} \bigskip
   \vskip\lettertopfil
   \ialign to\hsize{\strut ##\hfil\tabskip 0pt plus \hsize \cr #1 \crcr}
   \medskip\noindent\hskip\spskip}
\newskip\signatureskip	     \signatureskip=50pt
\def\signed#1{\par \penalty 9000 \bigskip \dt@pfalse
  \everycr={\noalign{\ifdt@p\vskip\signatureskip\global\dt@pfalse\fi}}
  \setbox0=\vbox{\singlespace \halign{\tabskip 0pt \strut ##\hfil\cr
   \noalign{\global\dt@ptrue}#1\crcr}}
  \line{\hskip 0.5\hsize minus 0.5\hsize \box0\hfil} \medskip }

\def\endletter{\ifnum\pagenumber=1 \vskip\letterbottomfil\supereject
\else \vfil\supereject \fi}
\newbox\letterb@x
\def\lettertext{\par\unvcopy\letterb@x\par}
\def\multiletter{\setbox\letterb@x=\vbox\bgroup
      \everypar{\vrule height 1\baselineskip depth 0pt width 0pt }
      \singlespace \topskip=\baselineskip }
\def\letterend{\par\egroup}
%
%%%%%%%%%%%%%%%%%%%%%%%%%%%%%%%%%%%%%%%%%%%%%%%%%%%%%%%%%%%%%%%%%%%%%%%
%
%   Here come macros for title pages.
%
\newskip\frontpageskip
\newtoks\pubtype
\newtoks\Pubnum
\newtoks\pubnum
\newif\ifp@bblock  \p@bblocktrue
\def\PH@SR@V{\doubl@true \baselineskip=24.1pt plus 0.2pt minus 0.1pt
	     \parskip= 3pt plus 2pt minus 1pt }
\def\PHYSREV{\paperstyle\PhysRevtrue\PH@SR@V}
%\def\titlepage{\FRONTPAGE\paperstyle\ifPhysRev\PH@SR@V\fi
%   \ifp@bblock\p@bblock\fi}
%
%AOKI
\newif\ifYUKAWA  \YUKAWAfalse
\def\YUKAWAmark{\hbox{\elevenmib
   Yukawa\hskip0.05cm Hall\hskip0.05cm Kyoto \hfill}}
\def\titlepage{\FRONTPAGE\paperstyle\ifPhysRev\PH@SR@V\fi
    \ifYUKAWA\null\vskip-1.70cm\YUKAWAmark\vskip0.6cm\fi
    \ifp@bblock\p@bblock\fi}
\def\nopubblock{\p@bblockfalse}
\def\endpage{\vfil\break}
\frontpageskip=1\medskipamount plus .5fil
%
%AOKI
\pubtype={ }
% 6/24/1986 The above was modified as three lines below by H.Mawatari
\newtoks\publevel
\publevel={Report}   % The alternatives are Internal and Preprint
%AOKI
\Pubnum={\the\pubnum}
% Nov.22/1987 IKEMORI
\pubnum={    }
%
%AOKI
\def\p@bblock{\begingroup \tabskip=\hsize minus \hsize
   \baselineskip=1.5\ht\strutbox \topspace-2\baselineskip
   \halign to\hsize{\strut ##\hfil\tabskip=0pt\crcr
   \the\Pubnum\cr \the\date\cr }\endgroup}
%
%\def\title#1{\vskip\frontpageskip \titlestyle{#1} \vskip\headskip }
%AOKI
\def\title#1{\vskip\frontpageskip\vfill
   {\fourteenbf\seventeenmin\titlestyle{#1}}\vskip\headskip\vfill }
\def\author#1{\vskip\frontpageskip\titlestyle{\twelvemin\twelvecp #1}\nobreak}
%\def\andauthor{\vskip\frontpageskip\centerline{and}\author}
%AOKI

%
%AOKI
\def\address#1{\par\kern 5pt \titlestyle{\twelvepoint\sl #1}}
\def\andaddress{\par\kern 5pt \centerline{\sl and} \address}
%
%\def\abstract{\vskip\frontpageskip\centerline{\fourteenrm ABSTRACT}
%	      \vskip\headskip }
%AOKI
\def\abstract#1{\vfill\vskip\frontpageskip
    \hbox{{\fourteencp Abstract}\hfill}\vskip\headskip#1\vfill\endpage}

%

%
%
%%%%%%%%%%%%%%%%%%%%%%%%%%%%%%%%%%%%%%%%%%%%%%%%%%%%%%%%%%%%%%%%%%%%%%%%
%   Miscellaneous macros
%

\def\\{\relax\ifmmode\backslash\else$\backslash$\fi}
\def\globaleqnumbers{\relax\if\equanumber<0\else\global\equanumber=-1\fi}

\def\journal#1&#2(#3){\unskip, \sl #1~\bf #2 \rm (19#3) }

\def\topspace{\hrule height 0pt depth 0pt \vskip}

\let\int=\intop		
\def\prop{\mathrel{{\mathchoice{\pr@p\scriptstyle}{\pr@p\scriptstyle}{
		\pr@p\scriptscriptstyle}{\pr@p\scriptscriptstyle} }}}
\def\pr@p#1{\setbox0=\hbox{$\cal #1 \char'103$}
   \hbox{$\cal #1 \char'117$\kern-.4\wd0\box0}}
\def\lsim{\mathrel{\mathpalette\@versim<}}
\def\gsim{\mathrel{\mathpalette\@versim>}}
\def\@versim#1#2{\lower0.2ex\vbox{\baselineskip\z@skip\lineskip\z@skip
  \lineskiplimit\z@\ialign{$\m@th#1\hfil##\hfil$\crcr#2\crcr\sim\crcr}}}
%
% % % % % % % % % % % % % % % % % % % % % % % % % % % % % % % % % % % %
%
%   Finally, some bug fixings.
%
\let\sec@nt=\sec
\def\sec{\relax\ifmmode\let\n@xt=\sec@nt\else\let\n@xt\section\fi\n@xt}
\def\obsolete#1{\message{Macro \string #1 is obsolete.}}
\def\firstsec#1{\obsolete\firstsec \section{#1}}
\def\firstsubsec#1{\obsolete\firstsubsec \subsection{#1}}
\def\thispage#1{\obsolete\thispage \global\pagenumber=#1\frontpagefalse}
\def\thischapter#1{\obsolete\thischapter \global\chapternumber=#1}
\def\nextequation#1{\obsolete\nextequation \global\equanumber=#1
   \ifnum\the\equanumber>0 \global\advance\equanumber by 1 \fi}
\def\BOXITEM{\afterassigment\B@XITEM\setbox0=}
\def\B@XITEM{\par\hangindent\wd0 \noindent\box0 }
%

%%%%%%%%%%%%%%%%%%%%%%%%%%%%%%%%%%%%%%%%%%%%%%%%%%%%%%%%%%%%%%%%%%%%%%%%
%   That's about it
%
\catcode`@=12 % at signs are no longer letters
\message{ by V.K.}
\relax

\def\yen{\hbox{Y\kern-0.75em =}}
%
%%%%%%%%%%%%%%%%%%%%%%%%%%%%%%%%%%%%%%%%%%%%%%%%%%%%%%
%% making \mib (math italic bold) available	    %%
%%		  by H. Mawatari, November 18, 1986 %%
%%%%%%%%%%%%%%%%%%%%%%%%%%%%%%%%%%%%%%%%%%%%%%%%%%%%%%
%
\catcode`@=11
%
%%%%%% start of net extension %%%%
%
%\font\seventeenmib=cmmib10 scaled\magstep3  \skewchar\seventeenmib='177
\font\fourteenmib=cmmib10 scaled\magstep2    \skewchar\fourteenmib='177
\font\twelvemib=cmmib10 scaled\magstep1	    \skewchar\twelvemib='177
\font\elevenmib=cmmib10 scaled\magstephalf   \skewchar\elevenmib='177
\font\tenmib=cmmib10			    \skewchar\tenmib='177
%\font\ninemib=cmmib9			    \skewchar\ninemib='177
%\font\sixmib=cmmib6			    \skewchar\sixmib='177
%
%\font\seventeenbsy=cmbsy10 scaled\magstep3   \skewchar\seventeenbsy='60
\font\fourteenbsy=cmbsy10 scaled\magstep2     \skewchar\fourteenbsy='60
\font\twelvebsy=cmbsy10 scaled\magstep1	      \skewchar\twelvebsy='60
\font\elevenbsy=cmbsy10 scaled\magstephalf    \skewchar\elevenbsy='60
\font\tenbsy=cmbsy10			      \skewchar\tenbsy='60
%\font\ninebsy=cmbsy9			       \skewchar\ninebsy='60
%\font\sixbsy=cmbsy6			       \skewchar\sixbsy='60
%
\newfam\mibfam
\def\samef@nt{\relax \ifcase\f@ntkey \rm \or\oldstyle \or\or
%	 \or\it \or\sl \or\bf \or\tt \or\caps \fi }
% 11/18/1986 modified as below by H.Mawatari
	 \or\it \or\sl \or\bf \or\tt \or\caps \or\mib \fi }
\def\fourteenpoint{\relax
    \textfont0=\fourteenrm	%@@\textfont0=\fourteenmin
    \scriptfont0=\tenrm
    \scriptscriptfont0=\sevenrm
     \def\rm{\fam0 \fourteenrm \f@ntkey=0 }\relax
    \textfont1=\fourteeni	    \scriptfont1=\teni
    \scriptscriptfont1=\seveni
     \def\oldstyle{\fam1 \fourteeni\f@ntkey=1 }\relax
    \textfont2=\fourteensy	    \scriptfont2=\tensy
    \scriptscriptfont2=\sevensy
    \textfont3=\fourteenex     \scriptfont3=\fourteenex
    \scriptscriptfont3=\fourteenex
    \def\it{\fam\itfam \fourteenit\f@ntkey=4 }\textfont\itfam=\fourteenit
    \def\sl{\fam\slfam \fourteensl\f@ntkey=5 }\textfont\slfam=\fourteensl
    \scriptfont\slfam=\tensl
    \def\bf{\fam\bffam \fourteenbf\f@ntkey=6 }\textfont\bffam=\fourteenbf
    \scriptfont\bffam=\tenbf	 \scriptscriptfont\bffam=\sevenbf
    \def\tt{\fam\ttfam \twelvett \f@ntkey=7 }\textfont\ttfam=\twelvett
    \h@big=11.9\p@{} \h@Big=16.1\p@{} \h@bigg=20.3\p@{} \h@Bigg=24.5\p@{}
    \def\caps{\fam\cpfam \twelvecp \f@ntkey=8 }\textfont\cpfam=\twelvecp
    \setbox\strutbox=\hbox{\vrule height 12pt depth 5pt width\z@}
% 11/18/1986 the following lines added by H.Mawatari
    \def\mib{\fam\mibfam \fourteenmib \f@ntkey=9 }
    \textfont\mibfam=\fourteenmib      \scriptfont\mibfam=\tenmib
    \scriptscriptfont\mibfam=\tenmib
% added up to here
    \samef@nt}
\def\twelvepoint{\relax
    \textfont0=\twelverm
    \scriptfont0=\ninerm
    \scriptscriptfont0=\sixrm
     \def\rm{\fam0 \twelverm \f@ntkey=0 }\relax
     \textfont1=\twelvei          \scriptfont1=\ninei
    \scriptscriptfont1=\sixi
     \def\oldstyle{\fam1 \twelvei\f@ntkey=1 }\relax
    \textfont2=\twelvesy	  \scriptfont2=\ninesy
    \scriptscriptfont2=\sixsy
    \textfont3=\twelveex	  \scriptfont3=\twelveex
    \scriptscriptfont3=\twelveex
    \def\it{\fam\itfam \twelveit \f@ntkey=4 }\textfont\itfam=\twelveit
    \def\sl{\fam\slfam \twelvesl \f@ntkey=5 }\textfont\slfam=\twelvesl
    \scriptfont\slfam=\ninesl
    \def\bf{\fam\bffam \twelvebf \f@ntkey=6 }\textfont\bffam=\twelvebf
    \scriptfont\bffam=\ninebf	  \scriptscriptfont\bffam=\sixbf
    \def\tt{\fam\ttfam \twelvett \f@ntkey=7 }\textfont\ttfam=\twelvett
    \h@big=10.2\p@{}
    \h@Big=13.8\p@{}
    \h@bigg=17.4\p@{}
    \h@Bigg=21.0\p@{}
    \def\caps{\fam\cpfam \twelvecp \f@ntkey=8 }\textfont\cpfam=\twelvecp
    \setbox\strutbox=\hbox{\vrule height 10pt depth 4pt width\z@}
% 11/18/1986 the following lines added by H.Mawatari
    \def\mib{\fam\mibfam \twelvemib \f@ntkey=9 }
    \textfont\mibfam=\twelvemib	    \scriptfont\mibfam=\tenmib
    \scriptscriptfont\mibfam=\tenmib
% added up to here
    \samef@nt}
\def\tenpoint{\relax
    \textfont0=\tenrm
    \scriptfont0=\sevenrm
    \scriptscriptfont0=\fiverm
    \def\rm{\fam0 \tenrm \f@ntkey=0 }\relax
    \textfont1=\teni	       \scriptfont1=\seveni
    \scriptscriptfont1=\fivei
    \def\oldstyle{\fam1 \teni \f@ntkey=1 }\relax
    \textfont2=\tensy	       \scriptfont2=\sevensy
    \scriptscriptfont2=\fivesy
    \textfont3=\tenex	       \scriptfont3=\tenex
    \scriptscriptfont3=\tenex
    \def\it{\fam\itfam \tenit \f@ntkey=4 }\textfont\itfam=\tenit
    \def\sl{\fam\slfam \tensl \f@ntkey=5 }\textfont\slfam=\tensl
    \def\bf{\fam\bffam \tenbf \f@ntkey=6 }\textfont\bffam=\tenbf
    \scriptfont\bffam=\sevenbf	   \scriptscriptfont\bffam=\fivebf
    \def\tt{\fam\ttfam \tentt \f@ntkey=7 }\textfont\ttfam=\tentt
    \def\caps{\fam\cpfam \tencp \f@ntkey=8 }\textfont\cpfam=\tencp
    \setbox\strutbox=\hbox{\vrule height 8.5pt depth 3.5pt width\z@}
% 11/18/1986 the following lines added by H.Mawatari
    \def\mib{\fam\mibfam \tenmib \f@ntkey=9 }
    \textfont\mibfam=\tenmib   \scriptfont\mibfam=\tenmib
    \scriptscriptfont\mibfam=\tenmib
% added up to here
    \samef@nt}
%
%%%%% end of net extension %%%%%
%
\Twelvepoint
\catcode`@=12
%
%%%%%%%%%%%%%%%%% end of \mib extension %%%%%%%%%%%%%%%%%%
%
%\everyjob{\input rifpdef\message{Good Luck}}
%
%\message{modified by K-I.AOKI}
\unnumberedchapters

\def\under#1{$\underline{\raise4pt{\hbox{#1}}}$}
\def\delbvec{\partial^{\raise3pt\hbox{$\!\!\!\!\leftrightarrow$}}}
\def\ymh{Yang-Mills-Higgs lagrangian$\;$}

\def\ws{Weinberg-Salam theory$\;$}
\def\dis{\displaystyle}
\def\={\,=\,}
\def\+{\,+\,}
\def\-{\,-\,}

\def\iii{I\hskip -0.4mm I\hskip -0.4mm I}
\def\cham{Chamseddine et al.}\titlepage
\title
{
Standard Model in Differential Geometry on  Discrete
Space $M_4\times Z_3$}
\author{Yoshitaka Okumura}
\address{\hskip -0.8cm
  Department of Applied Physics,
  Chubu University, Kasugai, Aichi, 487
}
\vskip 1cm
%\centerline{(October 20, 1993)}
\abstract{
Standard model is reconstructed using
the generalized differential calculus
extended on the discrete space $M_4\times Z_3$.
$Z_3$ is necessary for the inclusion of strong interaction.
Our starting point is the generalized gauge field expressed as
$A(x,y)=\!\sum_{i}a^\dagger_{i}(x,y){\bf d}a_i(x,y), (y=0,\pm)$,
where $a_i(x,y)$
is the square matrix valued function defined on $M_4\times Z_3$
and ${\bf d}=d+{\chi}$ is generalized
exterior derivative. We can construct the consistent algebra of $d_{\chi}$
with the introduction of the symmetry breaking function $M(y)$
%which is exterior derivative with respect to $Z_{3}$
and the spontaneous breakdown of gauge symmetry is coded in ${d_{\chi}}$.
%with  which introduces the matrix $M(y)$.
 The gauge field $A_\mu(x,y)$ and Higgs field $\Phi(x,y)$ are written
 in terms of $a_i(x,y)$ and $M(y)$, which might suggest $a_i(x,y)$ to be more
 fundamental object.
The unified picture of the gauge field and Higgs field as the generalized
connection in non-commutative geometry is realized.
Not only \ymh but also Dirac lagrangian, invariant against the gauge
transformation, are reproduced through the inner product between the
differential forms.
Two model constructions are presented, which are distinguished
in the particle assignment of Higgs field $\Phi(x,y)$.
%Within our framework, we can make the following predictions. The first model
% deduces the inequality $m_{\mathop{}_{H}} \leq \sqrt{2}m_{\mathop{}_{W}}$,
% whereas the second model leads to the interesting relation
% $m_{\mathop{}_{H}} =  2\sqrt{2}m_{\mathop{}_{W}}\sin \theta_{\mathop{}_{W}}$.
% This implies $m_{\mathop{}_{H}} =109.1GeV$ if
% $\sin^2 \theta_{\mathop{}_{W}}= 0.233$ and $m_{\mathop{}_{W}}=79.9GeV$
%  as the experimental values.
}

\section{
{
\hskip 5cm {\bf \S 1. Introduction}
}}
Higgs-Kibble mechanism is very important for the gauge theory
with the spontaneous breakdown of symmetry. \ws 1) based on this mechanism
 has elucidated many experimental mysteries of the electro-weak interactions
 since proposed in 1967.
 However, it seems to be rather artificial and Higgs
particle as the essential ingredient of this mechanism has not been detected
from now, which have persuaded one to investigate other alternatives 3)
such as
the technicolor model, Kaluza-Klein theory  and etc.. Such alternatives
, however, are by no means successful and satisfactory because of the extra
physical modes like the technicolor particles. \par
Recently, Connes 3) has proposed the new approach which enables us
to understand Higgs mechanism in non-commutative geometry on the discrete
space $M_4\times Z_2$. The unified picture of the gauge field and Higgs field
 as the generalized connection on the discrete space is realized.
 However, the mathematical settings of his theory are so
 difficult that one is hard to understand it. Several versions have been
 extended in line with Connes's original idea
 to make his approach more understandable.
Chamseddine, Felder and Frolich 4) provide more successful formalism
to be applicable to grand unified theories with the complicated breaking
 pattern of symmetry. They introduce the generalized Dirac operator and
 get the gauge invariant lagrangian by use of Clifford algebra. \par
Sitarz 5) investigated this problem in more familiar form with the ordinary
differential geometry. His approach is seemingly different to that of Connes.
He introduced the fifth one form basis $\chi$ and the corresponding exterior
derivative $d_\chi$ in addition to the ordinary one form basis $dx^\mu$ and
the exterior derivative $d$. His approach, however, seems to be difficult to
be applied to the gauge theory with rather complicated breaking pattern like
 the grand unified theory  \par
 K. Morita and the present author
6) developed this approach adopting the algebraic rules different from
that of Sitarz also on $M_4\times Z_2$. We introduced the matrix function
$M(y)$ ( $y$ is a variable in $Z_2$ and $y=\pm$ ) which is a factor
 to determine the scale and pattern of the spontaneous breakdown of
 gauge symmetry. The introduction of $M(y)$ makes the formalism more flexible.
 However, Model I in Ref.6) permits
 the appearance of gauge non-invariant term  in Higgs potential and
 we have to discard it by hand. This defect in our first paper 6)
  has been overcome by taking account of the auxiliary fields according to
  \cham 4)  and we could
  reproduce standard model in the completely gauge invariant way 7).
  However, the inclusion of the strong interaction into standard model  was not
  successful and the treatment of fermion sector was unsatisfactory.
  The purpose of this paper is to present the complete formalism of standard
model.
  The inclusion of strong interaction requires the discrete space
  $M_4\times Z_3$, not $M_4\times Z_2$. We can extend the formalism by use of
  only one form basis $\chi$ like in Ref.7) since the strong interaction
  does not break spontaneously. The generalization of the present formalism
  to $M_4\times Z_{\mathop{}_{N}}$ has been done by the present author 8)
  and $SU(5)$ grand unified theory is reconstructed therein.
This paper is divided into five sections and one Appendix. The next section
presents
the general framework based on the generalized differential calculus on
$M_4\times Z_3$. The generalized gauge field is defined there and a geometrical
picture for the unification of the gauge and Higgs fields is realized.
The third section provides the gauge invariant lagrangian for both boson
 and fermion sectors on the same footing. The fourth section is devoted to the
 reconstruction of the standard model. Two model constructions will be made.
 The last section is devoted for conclusions and discussions.
$SU(2)$ Higgs-Kibble model is treated in the non-commutative geometry
on $M_4\times Z_2$ in the Appendix.

\section{
{
\hskip 2.2cm
{\bf \S 2. The generalized connection on $M_4\times Z_3$ }
}}
The inclusion of strong interaction into standard model requires the discrete
space
$M_4 \times Z_3$ because the corresponding gauge symmetry is
$SU(3)_c\times SU(2)\times U(1)$.
Let $ A(x,y)$ be the generalized connection (gauge field) defined
on discrete space $M_4 \times Z_3$, where $M_4$ is the ordinary four
dimensional Minkowski space and $Z_3$ is the space with the discrete
variable $y\;(=0,\pm)$. According to \cham 4), we express $A(x,y)$ to be
$$
      A(x,y)\=\sum_{i}a^\dagger_{i}(x,y){\bf d}a_i(x,y),\eqno(1)
$$
where $a_i(x,y)$ the square-matrix-valued function
\footnote{*)}{
$a_i(x,0)$ is $3\times3$ ,
$a_i(x,+)$ is $2\times2$ and
$a_i(x,-)$ is $1\times1$ matrix.}
and ${\bf d}$ is the generalized exterior derivative defined as follows.
$$
       \eqalign{{\bf d}a_{i}(x,y)&\=(d + d_\chi)a_i(x,y)\cr
     da_i(x,y) &\= \partial_\mu a_i(x,y)dx^\mu,\hskip 1cm
     \partial_\mu \equiv {\partial \over \partial x^\mu},\cr
     d_\chi a_i(x,y) &\= [-a_i(x,y)M(y) + M(y)a_i(x,-y)]\chi, \cr}
        \eqno(2)
$$
Here $dx^\mu$ is
ordinary one form basis, taken to be dimensionless, in $M_4$, and $\chi$
is the one form basis in the discrete space $Z_3$ assumed
to be also dimensionless,
 and we assume $\chi^\dagger\=-\chi$.
We have introduced $x$-independent matrix $M(y)$ whose hermitian conjugation is
given by $M(y)^\dagger=M(-y)$. The matrix $M(y)$ turns out to determine the
scale and pattern of the spontaneous breakdown of the gauge symmetry. We take
$M(0)=0$ because $SU(3)_c$ as the symmetry of strong interaction
is not spontaneously broken. For this reason, one-form basis in $Z_3$ is
only $\chi$.
\par
\noindent
We further define the calculation rules as follows.
$$
      \eqalign{ a_i(x,y)dx^\mu &= dx^\mu a_i(x,y),\cr
              f(x,y)\chi &= \chi f(x,-y), \cr
               d_\chi \chi&= 0,   \cr
              d_\chi M(y) &= M(y)M(-y)\chi,
                          \cr}
                          \eqno(3)
$$
where $f(x,y)$ denotes $a_i(x,y)$ or $M(y)$.
The second equation in Eq.(3) expresses the non-commutativity of the geometry
under consideration. We have one more assumption that
\footnote{*)}{
For example, we take the following manipulation.
$$
\eqalign{
d_\chi(a_i(x,y)M(y)&b_i(x,-y)) = (d_\chi a_i(x,y))M(y)b_i(x,-y) \cr
      &   + a_i(x,y)(d_\chi M(y))b_i(x,-y)
       - a_i(x,y)M(y)(d_\chi b_i(x,-y)).\cr}
$$
}
$$
              d_\chi (p(x,y)q(x,y)) = (d_\chi p(x,y))q(x,y)+
                          (-1)^{r} p(x,y)(d_\chi q(x,y)),
$$
where $p(x,y)$ and $q(x,y)$ consist of the products of $a_i(x,y)$ and $M(y)$
like in the footnote and $r$ is the number of $M(y)$ in $p(x,y)$.
According to these calculation rules we can prove  Leibniz rule for the product
of square matrix functions as
$$
               d_\chi (a_{i}(x,y)b_{i}(x,y)) \= (d_\chi a_{i}(x,y))b(x,y)+
                   a_{i}(x,y)(d_\chi b_{i}(x,y)).
$$
With these considerations we easily prove the nilpotency of $\chi$
$$
    d_\chi^2a_i(x,y)=0, \hskip 1.5cm d_\chi^2M(y)=0, \eqno(4)
$$
which leads to the nilpotency  of ${\bf d}$ in the follows.
$$
{\bf d}^2 a_i(x,y)=0, \hskip 1.5cm
 {\bf d}^2 M(y)=0,
\eqno(5)
$$
where $dx^\mu\wedge dx^\nu=-dx^\nu\wedge dx^\mu$
and $dx^\mu\wedge \chi_m=-\chi_m\wedge dx^\mu$
are assumed reasonable in the ordinary differential calculus.
The proof of  ${\bf d}^2 p(x,y)\=0$ is also easy, but we quote only Eq.(5)
in this paper. Nilpotency ${\bf d}^2 = 0$ is very important for the gauge
invariant formulation.
\par
Let us rewrite $A(x,y)$ in Eq.(1) by use of Eq.(2).
$$
     \eqalign{A(x,y) &\= \sum_{i}a_{i}^\dagger(x,y)
       [\partial_\mu a_{i}(x,y)dx^\mu +
                (-a_i(x,y)M(y) + M(y)a_i(x,-y))\chi] \cr
         &\= \sum_{i}a_{i}^\dagger(x,y)\partial_\mu a_{i}(x,y)dx^\mu \cr
        &\hskip 1.5cm + \sum_{i}a_{i}^\dagger(x,y)(-a_i(x,y)M(y) +
M(y)a_i(x,-y))\chi.\cr}
  \eqno(6)
$$
{}From this equation we can define the ordinary gauge field $A_\mu (x,y)$
and Higgs field $\Phi(x,y)$ as follows.
$$
  \eqalign{
    A_\mu(x,y) &= \sum_{i}a_{i}^\dagger(x,y)\partial_\mu a_{i}(x,y), \cr
\Phi(x,y) &= \sum_{i}a_{i}^\dagger(x,y)(-a_i(x,y)M(y) + M(y)a_i(x,-y)).\cr
}  \eqno(7)
$$
Here we require according to \cham 4)
$$
\sum_{i}a_{i}^\dagger(x,y)a_{i}(x,y)\=1,  \eqno (8)
$$
without loss of generality, which leads to
$$
   A_\mu^\dagger(x,y) \= -A_\mu (x,y), \hskip 2cm
   \Phi^\dagger (x,y) \= \Phi(x,-y).  \eqno(9)
$$
Eqs.(7) and (8) make us perceive the necessity of the sum about $i$
in Eq.(1) in order for $A_\mu(x,y)$ not to become the pure gauge field.
The generalized connection $A(x,y)$ is written in terms of
$A_\mu (x,y)$ and $\Phi(x,y)$ as
$$
    A(x,y) \= A_\mu(x,y)dx^\mu + \Phi(x,y)\chi.  \eqno(10)
$$
Eq.(10) realizes the unified picture of the gauge field and Higgs field as
the generalized connection in the geometry on the discrete space
$M_4\times Z_3$.
Since the fifth dimension is discrete, there arises no additional
physical degrees of freedom as in the Kaluza-Klein theory.
\par
The generalized field strength  ${F}(x,y)$ is defined in terms of $A(x,y)$ as
$$
    { F}(x,y) \= {\bf d}A(x,y) + A(x,y)\wedge A(x,y). \eqno(11)
$$
In order to explicit expression of ${F}(x,y)$, we have to calculate
${\bf d}A(x,y)$ from Eq.(1). Owing to the nilpotency ${\bf d}^2 = 0$ in Eq.(5)
we find
$$
   \eqalign{
    {\bf d}A(x,y) &\= {\bf d}(\sum_{i}a^\dagger_{i}(x,y){\bf d}a_i(x,y))\cr
             & \= \sum_{i}{\bf d}a^\dagger_{i}(x,y) \wedge {\bf d}a_i(x,y).\cr}
   \eqno(12)
$$
Inserting  Eq.(2) into Eq.(12) and using Eqs.(7) and (8), we obtain
$$
   \eqalign{
    {\bf d}A(x,y) &\= \partial_\mu A_\nu(x,y) dx^\mu \wedge dx^\nu \cr
       &\hskip 0.5cm + [\partial_\mu \Phi(x,y) + A_\mu(x,y)M(y) - M(y)A_\mu(x
             ,-y)]dx^\mu \wedge \chi \cr
      &\hskip 0.5cm +[\Phi(x,y)M(-y) + M(y)\Phi(x,-y) \cr
      &\hskip 1cm +M(y)M(-y) - \sum_{i}a_{i}^\dagger(x,y)M(y)M(-y)a_{i}(x,y)]
        \chi \wedge \chi. \cr} \eqno(13)
$$
The auxiliary field $Y(x,y)$ is denoted as
$$
  Y(x,y)\=\sum_{i}a_{i}^\dagger(x,y)M(y)M(-y)a_{i}(x,y), \eqno(14)
$$
which plays an important role to insure the gauge invariance of
\ymh as shown later.
  $A(x,y) \wedge A(x,y)$ in Eq.(11) is calculated by use of Eq.(9) as
$$
\eqalign{
    A(x,y) \wedge A(x,y) &\= A_\mu(x,y)A_\nu(x,y)dx^\mu \wedge dx^\nu \cr
          &\hskip 1cm
          + (A_\mu(x,y)\Phi(x,y) - \Phi(x,y)A_\mu(x,-y))dx^\mu \wedge \chi \cr
        &\hskip 1cm
          + \Phi(x,y)\Phi(x,-y)\chi \wedge \chi. \cr} \eqno(15)
$$
{}From Eqs.(13) and (15) the generalized field strength ${F}(x,y)$ is written
as
$$
 {F}(x,y) = { 1 \over 2}F_{\mu\nu}(x,y)dx^\mu \wedge dx^\nu +
               D_\mu\Phi(x,y)dx^\mu \wedge \chi + V(x,y)\chi \wedge \chi,
               \eqno(16)
$$
where
$$
  \eqalign{
  F_{\mu\nu}(x,y)&=\partial_\mu A_\nu (x,y) - \partial_\nu A_\mu (x,y)
               + [A_\mu(x,y), A_\mu(x,y)],\cr
  D_\mu\Phi(x,y)&=\partial_\mu \Phi(x,y) - (M(y) + \Phi(x,y))A_\mu(x,-y)\cr
         & \hskip 1.5cm + A_\mu(x,y)(M(y) + \Phi(x,y))\cr
  V(x,y)&= (\Phi(x,y) + M(y))(\Phi(x,-y) + M(-y)) -
             Y(x,y).\cr}
     \eqno(17)
$$
If we put $H(x,y) = \Phi(x,y) + M(y)$,
the second and third equations in Eq.(17) is rewritten as
$$
  \eqalign{
  D_\mu\Phi(x,y)&=\partial_\mu H(x,y) + A_\mu(x,y)H(x,y) - H(x,y)A_\mu(x,-y)\cr
  V(x,y)&= H(x,y)H(x,-y) -   Y(x,y),\cr}
  \eqno(18)
$$
which indicates that $H(x,y)$ is the un-shifted Higgs field with the vacuum
expectation value $M(y)$ whereas $\Phi(x,y)$ represents the physical field
with the vanishing vacuum expectation value. This is the meaning that $M(y)$
 determines the scale and pattern of the spontaneous breakdown of gauge
 symmetry.
\par
In order to construct gauge invariant lagrangian
we have to investigate the gauge transformation property.
The gauge transformation is defined by
$$
 A^{g}(x,y)=g^{-1}(x,y)A(x,y)g(x,y)+g^{-1}(x,y){\bf d}g(x,y),
\eqno(19)
$$
where  $g(x,y)$ is the gauge function with respect to the corresponding unitary
group. We take the operation of ${\bf d}=d+\chi$ on $g(x,y)$ as
$$
       \eqalign{{\bf d}g(x,y)&\=(d + d_\chi)g(x,y)\cr
              &= \partial_\mu g(x,y)dx^\mu
             + [-a_i(x,y)M(y) + M(y)a_i(x,-y)]\chi, \cr}
$$
which together with Eq.(19) defines the gauge transformation
law for $A_\mu(x,y)$ and $\Phi(x,y)$,
$$
\eqalign{
   A_\mu^{g}(x,y) &\= g^{-1}(x,y)\partial_\mu g(x,y) +
                      g^{-1}(x,y)A_\mu(x,y)g(x,y) \cr
\Phi^g(x,y)&\=g^{-1}(x,y)\Phi(x,y)g(x,-y)+g^{-1}(x,y)(
M(y)g(x,-y)-g(x,y)M(y)).\cr}
\eqno(20)
$$
For the unphysical Higgs field $H(x,y)$, we have
$$
H^{g}(x,y)=g^{-1}(x,y)H(x,y)g(x,-y).
\eqno(21)
$$
This is the usual gauge transformation law of Higgs field when
the spontaneous breakdown of symmetry is not still switched on.
The gauge transformation Eqs.(19) and (20) require the gauge transformation
for the fundamental field $a_i(x,y)$ to be
$$
      a^{g}_{i}(x,y) \= a_{i}(x,y)g(x,y). \eqno(22)
$$
Then, the gauge transformation of field strength is then subject to
$$
{ F}^{g}(x,y)=g^{-1}(x,y){ F}(x,y)g(x,y),
\eqno(23)
$$
which shows $F(x,y)$ gauge covariant.
In components, Eq.(23) reads
$$\eqalign{
F_{\mu\nu}^g(x,y)&=g^{-1}(x,y)F_{\mu\nu}(x,y)g(x,y),\cr
D_\mu^g\Phi(x,y)&=g^{-1}(x,y)D_\mu\Phi(x,y)g(x,-y),\cr
V^g(x,y)&=g^{-1}(x,y)V(x,y)g(x,y).\cr
}
\eqno(24)
$$
In order for the third equation in Eq.(24) to be consistent,
Eq.(22) is inevitable.
With these gauge transformations, the gauge invariant Yang-Mills-Higgs
lagrangian is obtainable. \par
\section{
\hskip 3.5cm
{\bf  \S 3. Gauge invariant lagrangian}
}
In order to get the \ymh from the field strength ${F}(x,y)$ which is the
differential two form, we have to decide the metric structure
of the discrete space $M_4 \times Z_3$.
The ordinary Minkowski
metric in $M_4$ is to be supplemented by
additional assumption that
the fifth direction is orthogonal to $M_4$
and the metric along the fifth direction is left arbitrary:
$$
\eqalign{
<dx^\mu, dx^\nu>&=g^{\mu\nu},\quad
g^{\mu\nu}={\rm diag}(1,-1,-1,-1),\cr
<\chi, dx^\mu>&=0,\cr
<\chi, \chi>&=-\alpha^2.\cr
}
\eqno(25)
$$
The inner products of two-forms
are taken to be
$$
\eqalign{
<dx^\mu \wedge dx^\nu,
dx^\rho \wedge dx^\sigma>&=g^{\mu\rho}g^{\nu\sigma}-
g^{\mu\sigma}g^{\nu\rho},\cr
<dx^\mu \wedge \chi,
dx^\nu \wedge \chi>&=-\alpha^2g^{\mu\nu},\cr
<\chi \wedge \chi,
\chi \wedge \chi>&=\beta^4,\cr
}
\eqno(26)
$$
while other inner products among the basis two-forms vanish.
It should be noticed that $\beta$ is not necessarily equal to $\alpha$
because of the non-commutative property of geometry under consideration.
\par
\noindent
\centerline{\bf Yang-Mills-Higgs sector}
\par
According to these metric structures and Eq.(16) we can give
the formula for gauge-invariant \ymh
$$
\eqalign{
{\cal L}_{\mathop{}_{YMH}}(x)=&-{\rm tr}\sum_{y=0,\pm}{1 \over g_{y}^2}
<{F}(x,y),
{F}(x,y)>\cr
=&-{\rm tr}\sum_{y=0,\pm}{1\over 2g^2_y}
F_{\mu\nu}^{\dag}(x,y)F^{\mu\nu}(x,y)\cr
+&{\rm tr}\sum_{y=0,\pm}{\alpha^2\over g_{y}^2}(D_\mu\Phi(x,y))^{\dag}
D^\mu\Phi(x,y)
-{\rm tr}\sum_{y=0,\pm}{\beta^4\over g_{y}^2}V^{\dag}(x,y)V(x,y),\cr
}
\eqno(27)
$$
where $g_y$ is the gauge coupling constant
and
tr denotes the trace over internal symmetry matrices. In Ref.6) we took $g_y$
to be common
value for $y=\pm$. However, in this paper we take $g_y$ to be different values
for $y=0,\pm$ to discuss  the inclusion of strong interaction.
%and possible quantum corrections in sec.V
It should be remarked that
Eq.(27)is \ymh with the same number of parameters as in the ordinary \ws.
We can not predict anything about Weinberg angle or Higgs mass
in contrast to the situation in the non-commutative geometry approach.
This is because $\beta$ is not necessarily equal to $\alpha$.
It may be interesting to investigate under what conditions $\beta$ is equal to
$\alpha$. It probably determines the non-commutativity of geometry under
consideration.
%, there exists a gauge invariant term with dimension 4
%, which indicated by Sitarz [5]. It is expressed as $tr\{g_{\alpha\beta}
%F^{\alpha\beta}\}^2=tr\{V(x,y)\}^2$. We call such a term  the anomalous term
% in our formalism.
\par
\noindent
\centerline{\bf fermionic sector}
\par
We propose the general framework to construct
the gauge invariant Dirac lagrangian of the fermionic sector.
\cham and others usually give the Dirac lagrangian
 by use of Clifford algebra. We want to get it in the same way as \ymh
 in Eq.(27) is introduced by taking the inner product of the differential
  forms.
\par
Remarking the equation ${\bf d}+A(x,y)$, we define
$$
{\cal D}\psi(x,y)=({\bf d}+ A^f(x,y))\psi(x,y),
\eqno(28)
$$
which we call the covariant spinor one-form.
Here, $A^f(x,y)$ corresponds to the differential representation
for $\psi(x,y)$
such that  $ A^f(x,y)=A_\mu^f(x,y)dx^\mu +\Phi(x,y)\chi$.
 Since the role of $d_\chi$ makes the shift  $\Phi(x,y)
\rightarrow \Phi(x,y)+M(y)$ as shown previously, we define
$$
d_\chi \psi(x,y)=M(y)\psi(x,-y)\chi.
\eqno(29)
$$
${\bf d}+ A^f(x,y)$ is Lorentz invariant, and so
${\cal D}\psi(x,y)$ is transformed as a spinor just like $\psi(x,y)$ against
Lorentz transformation.
Let $\rho$ be the representation for $\psi(x,y)$
so that the gauge transformation is expressed as
$$
    \psi^g(x,y)=\rho(g^{-1}(x,y))\psi(x,y).
$$
As $A^f_\mu(x,y)$ is Lee algebra for the fermion representation of gauge group
under consideration, it is transformed as
$$
 (A^f_\mu(x,y))^g=\rho(g^{-1}(x,y)) A^f_\mu(x,y)\rho(g(x,y))
+\rho(g^{-1}(x,y)){\partial_\mu}\rho(g(x,y)).
$$
In addition, we have to assign $\psi(x,y)$ in the model building
in order that the following relation is satisfied.
$$
      H^g(x,y)\psi^g(x,-y)=\rho(g^{-1}(x,y))H(x,y)\psi(x,-y),
$$
which is sufficiently taken into account in the following section.\hfill\break
{}From these, ${\cal D}\psi(x,y)$ is shown to be the gauge covariant;
$$
 {\cal D}^{g}\psi^{g}(x,y) = \rho(g^{-1}(x,y)){\cal D}\psi(x,y).
 \eqno(30)
$$
\par
${\cal D}\psi(x,y)$ just introduced above is the gauge covariant spinor one
form.
In order to give the gauge invariant Dirac lagrangian by taking
the inner product of differential form,
we have to introduce the corresponding spinor one form.
For this purpose we consider the following spinor one form.
$$
{\tilde {\cal D}}\psi(x,y)=\gamma_\mu \psi(x,y)dx^\mu+ic\psi(x,y)\chi,
\eqno(31)
$$
where  $c$ is a constant related with the Yukawa coupling between
Higgs particle and fermion.
${\tilde {\cal D}}\psi(x,y)$ is evidently transformed against  Lorentz
transformation  just as $\;\psi(x,y)$ and also gauge covariant. \par
Dirac lagrangian is computed by the inner product
$$\eqalign{
{\cal L}_{\mathop{}_{D}}(x,y)=
      &i<{\tilde {\cal D}}\psi(x,y),{\cal D}\psi(x,y)>\cr
=&i[{\bar\psi}(x,y)\gamma^\mu(\partial_\mu+A_\mu^f(x,y))\psi(x,y)\cr
&+i{\bar\psi}(x,y)c\alpha^2(M(y)+\Phi(x,y))\psi(x,-y)],\cr
}
\eqno(32)
$$
where we have defined the inner product for spinor one$-$form, by noting
 Eq.(25),
$$\eqalign{
<A(x,y)dx^\mu, B(x,y')dx^\nu>&=\bar{A}(x,y)B(x,y')g^{\mu\nu},\cr
<A(x,y)\chi, B(x,y')\chi>&=-\bar{A}(x,y)B(x,y')\alpha^2,\cr
}
\eqno(33)
$$
with vanishing other inner products and $\bar{A}(x,y)=A^{\dag}(x,y)\gamma^0$.
Yukawa interaction of Higgs particle to fermion is given
by the last term in Eq.(32).
$$
{\cal L}_{\mathop{}_{{\rm Yukawa}}}(x,y)
          =-g_{\mathop{}_{Y}}{\bar\psi}(x,y)H(x,y)\psi(x,-y), \eqno(34)
$$
which is evidently gauge invariant.
\par
The total Dirac lagrangian is the sum over $y=0,\pm$:
 \footnote{*)}{
We can also define the Dirac lagrangian by
$$
{\cal L}_D(x,y)=-i<{\cal D}(x,y)\psi(x,y),
{\tilde {\cal D}}(x,y)\psi(x,y)>
$$
which is the same as Eq.(32) except for the total divergence
and we assign $\;\psi(x,0)=0$ as shown later.}
$$
{\cal L}_{\mathop{}_{D}}(x)
        =\sum_{y=0,\pm}{\cal L}_{\mathop{}_{D}}(x,y),
\eqno(35)
$$
which is gauge and Lorentz invariant and hermitian.
Eqs.(27) and (35) play an important role to reconstruct standard model.
\par

\section{
%\centerline
{\hskip 3cm \bf \S 4 Reconstruction of  standard model}
}
In this section we present two model constructions
using the formalism developed in the previous section.
They are
designed so as to reproduce standard model with constraints from the
non-commutative
geometry.
\par
\noindent
\leftline{\bf Model I}
\par
Let us begin with  Yang-Mills-Higgs sector according to Eq.(27).
\par
\noindent
\centerline{\bf  Yang-Mills-Higgs sector}
\par
The first model we shall consider is based on the following identification
for $A(x,y), \Phi(x,y),$ and $M(y)\;(y=0,+,-)$ with obvious notations:
$$
\eqalign{
    A_\mu(x,0)&= -{i\over 2}\sum_{a=1}^8 \lambda^a G_\mu^a(x),\cr
    \Phi(x,0)&=0, \hskip 1cm M(0)=0 \cr
    g(x,0)&=g_s(x),\hskip 1cm g_s(x)\in SU(3)_c,\cr}
    \eqno(36a)
$$
where $G_\mu^a(x)$ represents gluon field, and the second equation means
$SU(3)_c$ gauge symmetry not to break and $\lambda^a(a=1,2\cdots 8)$
are Gell-Mann zweig matrices.
$$\eqalign{
A_\mu(x,+)&=-{i \over 2}\sum_{i=1}^3\tau^i A^i_\mu(x)-{i \over
2}a\tau^0B_\mu(x),\cr
\Phi(x,+)&=\Phi(x)=
\left(
\matrix{
\phi_+(x)\cr
\phi_0(x)\cr
}\right), \hskip 1cm
M(+)=
\left(
\matrix{
0\cr
\mu\cr
}\right),\cr
 g(x,+)&=e^{-ia\alpha(x)}g(x),\;e^{-ia\alpha(x)}\in U(1),\;
g(x)\in SU(2),\cr
}
\eqno(36b)
$$
where $A_\mu^i(x)$ and $B_\mu(x)$ denotes $SU(2)$ and $U(1)$
gauge fields, respectively. $B!!!!(J $\tau^i(i=1,2,3)$ are Pauli matrices
and $\tau^0$ is $2\times 2$ unit matrix..
%\vskip -0.2cm
$$\eqalign{
A_\mu(x,-)&=-{i \over 2}bB_\mu(x),\cr
\Phi(x,-)&=\Phi^{\dag}(x), \hskip 1cm
M(-)=(0,\;\; \mu)=M^{\dag}(+),\cr
g(x,-)&=e^{-ib\alpha(x)}\in U(1).\cr
}
\eqno(36c)
$$
$M(y)$ must be chosen to give the correct symmetry breakdown.
It should be noticed that there are free parameters $a,b$ in
Eqs.(36a) and (36b).
This implies that $U(2)$
on the upper sheet $y=+$
is not orthogonal
to $U(1)$ on the lower sheet $y=-$,
but the gauge group is only $SU(2) \times U(1)$.
\par
With above identifications the generalized gauge field $F(x,y)$
is expressed as
$$
\eqalign{
 F(x,0)&= -{i \over 4}\sum_a \lambda^a G_{\mu\nu}^a(x) dx^\mu \wedge dx^\nu,\cr
 F(x,+)&= {i\over 4}[\,- \sum_i\tau^i F_{\mu\nu}^i(x)
          -a\tau^0 B_{\mu\nu}(x)\, ]dx^\mu \wedge dx^\nu\cr
               + & D_\mu H(x)dx^\mu \wedge \chi
         +[\,H(x)H^\dagger(x)-Y(x,+)\,]\chi\wedge\chi,\cr
  F(x,-)&= -b{i\over 4}B_{\mu\nu}(x)dx^\mu \wedge dx^\nu \cr
            &+(\,D_\mu H(x)\,)^\dagger dx^\mu \wedge \chi
      + [\,H^\dagger(x)H(x)-Y(x,-)\,]\chi\wedge\chi,\cr
   } \eqno(37)
$$
where
$$
\eqalign{
      G_{\mu\nu}^a(x)=&\partial_\mu G_\nu^a(x)-\partial_\nu G_\mu^a(x)
         +f^{abc}G_\mu^b(x)G_\nu^c(x),  \cr
      F_{\mu\nu}^i(x)=&\partial_\mu A_\nu^i(x)-\partial_\nu A_\mu^i(x)
         +\epsilon^{ijk}A_\mu^j(x)A_\nu^k(x),  \cr
      B_{\mu\nu}(x)=&\partial_\mu B_\nu(x)-\partial_\nu B_\mu(x),\cr
      H(x)=&\Phi(x)+M(+).\cr
      D^\mu H(x)=&[\,\partial_\mu-{i\over 2}\,(\sum_i\tau^iA_\mu^i(x)
          +(a-b)\,\tau^0\,B_\mu(x)\,)\,]\,H(x). \cr
      } \eqno(38)
$$
By inserting Eq.(37) into Eq.(27), we immediately get
$$\eqalign{
{\cal L}_{{\mathop{}_{YMH}}}=&-\sum_{y=0,\pm}{1\over g_y^2}<F(x,y), F(x,y)>\cr
=&-{1\over 4g_s^2}\sum_a G_{\mu\nu}^a(x)G^{a\mu\nu}(x)\cr
&-{1\over 4g_+^2}
\sum_i F^i_{\mu\nu}(x)\cdot F^{i\mu\nu}(x)
-{1\over4}
({a^2\over g_+^2}+{b^2\over 2g_-^2})B_{\mu\nu}(x)B^{\mu\nu}(x))\cr
&+({1\over g_+^2}+{1\over g_-^2})\alpha^2(D_\mu H(x))^{\dag}D^\mu H(x)\cr
&-{\beta^4\over g_+^2}\,{\rm tr}\,(H(x)H^{\dag}(x)-Y(x,+))^2
-{\beta^4\over g_-^2}(H^{\dag}(x)H(x)-Y(x,-))^2.\cr
}
\eqno(39)
$$
Let us investigate the auxiliary fields $Y(x,+)$ and $Y(x,-)$ in Eq.(39)
from Eq.(14).
$$
  \eqalign{
   Y(x,+) &= \sum_{i}a_i^\dagger(x,+)M(+)M(-)a_i(x,+) \cr
          &= \sum_{i}a_i^\dagger(x,+)
          \left(\matrix{0 & 0 \cr
                        0 & \mu^2 \cr} \right) a_i(x,+).\cr} \eqno(40)
$$
This field can not be expressed in terms of gauge or Higgs fields
and also not a constant so that $Y(x,+)$ is independent.
Thus, the potential term containing this auxiliary field in Eq.(39)
is eliminated owing to the equation of motion.
In Ref.6), the term corresponding to this gauge non-invariant term
is discarded by hand, which is justified here.
$$
  \eqalign{
   Y(x,-) &= \sum_{i}a_i^\dagger(x,-)M(-)M(+)a_i(x,-) \cr
          &= \sum_{i}a_i^\dagger(x,-) \mu^2 a_i(x,-)= \mu^2,\cr}\eqno(41)
$$
which is a constant and yields the meaningful Higgs potential term
in the lagrangian Eq.(39).
\par
After rescaling
$$
\eqalign{
G_\mu(x) &\rightarrow g_s G_\mu(x),\cr
A_\mu^i(x)  &\rightarrow g_+ A_\mu^i(x),\cr
B_\mu(x) &\rightarrow {g_+g_-\over
      \sqrt{a^2g_-^2+{\dis{b^2\over 2}}g_-^2}}B_\mu(x),\cr
H(x) &\rightarrow{g_+g_-\over \sqrt{g_+^2+g_-^2}\alpha}H(x)
         =g_{\mathop{}_{H}}H(x),\cr
}  \eqno(42)
$$
we find the standard \ymh in standard model.
$$\eqalign{
{\cal L}_{{\mathop{}_{YMH}}}=&
-{1\over 4}\sum_a G_{\mu\nu}^a(x)\cdot G^{a\mu\nu}(x)\cr
  &-{1\over 4}(\sum_i F_{\mu\nu}^i(x)\cdot F^{i\mu\nu}(x)
  + B_{\mu\nu}(x)\cdot B^{\mu\nu}(x)\,) \cr
&+(D_\mu H(x))^{\dag}(D^\mu H(x))
-\lambda(\,H^{\dag}(x)\,H(x)-\mu'^2)^2,\cr
}
\eqno(43)
$$
where
$$
\eqalign{
G_{\mu\nu}^a (x) &= \partial_\mu G_\nu^a(x)-\partial_\nu G_\mu^a(x)
       +g_sf^{abc}G_\mu^b G_(x)\nu^c(x), \cr
F_{\mu\nu}^i (x) &= \partial_\mu A_\nu^i(x)-\partial_\nu A_\mu^i(x)
       +g\,\epsilon^{ijk}A_\mu^j A_(x)\nu^j(x), \cr
D_\mu H(x)&=[\partial_\mu-{i \over 2}(g\,\sum_i \tau^i\cdot  A^i_\mu(x)
+g'\tau_0B_\mu(x))]H(x),\cr}
\eqno(44)
$$
with $g_+=g,\;\dis{g^{'}={(a-b)g_+g_-\over \sqrt{a^2g_-^2+{\dis{b^2\over 2}}
g_+^2}}}\,$,  $\dis{\lambda={\beta^4g_+^{4}g_-^{2}
\over \alpha^4(g_+^2+g_-^2)^2}\,}$, and
 $\;\mu'\,=\dis{{\sqrt{g_+^2+g_-^2}\alpha\mu \over g_+g_-}}$.
  Eq.(43) expresses \ymh of the gauge theory with the symmetry
  $SU(3)_c\times SU(2)_{\mathop{}_{L}} \times U(1)$
 spontaneously broken to
 $SU(3)_c\times U(1)_{\rm em}$.
% \hfill\break
Denoting W,Z gauge bosons and photon by
$$
\eqalign{
W_\mu&={1\over \sqrt{2}}(A_\mu^1-iA_\mu^2)\,, \hskip 3.5cm
W_\mu^{\dag}={1\over \sqrt{2}}(A_\mu^1+iA_\mu^2), \cr
Z_\mu&={1\over {\sqrt{g^2+g^{'2}}}}(-g\,A_\mu^3+g'B_\mu), \hskip 1.5cm
A_\mu={1\over {\sqrt{g^2+g^{'2}}}}(g\,A_\mu^3+g'B_\mu), \cr
}
$$
we get the gauge boson mass term to be
$$
{\cal L}_{{\rm gauge}\;{\rm boson}\;{\rm mass}}
     =m_{\mathop{}_{W}}^2W_\mu^{\dag}W^\mu
     +{1\over 2}m_{ \mathop{}_{Z}}^2Z_\mu Z^\mu.\eqno(45)
$$
The gauge boson masses are explicitly given as
$$
 \eqalign{
 m_{ \mathop{}_{W}}&=\sqrt{{{1 + \delta^2} \over 2}}\alpha\mu,\cr
 m_{ \mathop{}_{Z}}&=\sqrt{{1+\delta^2 \over 2}(
 1+{2(a^-b)^2\over {2a^2+\delta^{2}b^2}})}\alpha\mu.\cr}\eqno(46)
$$
where $\dis{\delta={g_+\over g_-}}$.
Weinberg angle is determined by two parameters $r={\dis {b\over a}}$
and $\delta$:
$$
\sin^2{\theta_{ \mathop{}_{W}}}=1-{m_{ \mathop{}_{W}}^2\over
m_{ \mathop{}_{Z}}^2}={2(1-r)^2\over
{4-4r+(2+\delta^2)r^2}}.
\eqno(47)
$$
One usually impose the condition [3] ${\rm tr}A_\mu(x,+)={\rm tr}A_\mu(x,-)$
, which yields $r=1$ and so $\sin^2\theta_{\mathop{}_{W}}=\dis{{1 \over
2+2\delta^2}}\;$.
Furthermore, putting  $\delta=1$ as usual,
$\sin^2\theta_{\mathop{}_{W}}=\dis{{1 \over 4}}\;$, which value is
predicted by the original Connes's work. If $\beta=\sqrt{\eta}\alpha$, the
Higgs mass
is given from Eq.(45) by
$$
   m_{ \mathop{}_{H}} = { 2\eta\delta \over \sqrt{1+\delta^2}}\alpha\mu,
$$
which is related to the $W$ gauge boson mass via
$$
m_{ \mathop{}_{H}}={2\sqrt{2}\eta\delta \over 1+\delta^2}m_{ \mathop{}_{W}}.
\eqno(48)
$$
%If we put $\eta=1$ as usual, Eq.(48) leads to the inequality
% $m_{ \mathop{}_{H}} \leq {\sqrt 2}m_{ \mathop{}_{W}}$.
%It should be noted that this inequality has nothing to do with $r$.
If we put $\delta=1, \; \eta=1$ as usual, $m_{ \mathop{}_{H}}={\sqrt 2}m_{
\mathop{}_{W}}$,
which is typical 3) in non-commutative geometry.
$\alpha=\beta\;(\eta=1)$ is natural in the commutative geometry.
It is interesting to investigate whether there is a symmetry corresponding to
$\alpha=\beta$ in the non-commutative geometry on the discrete space $Z_3$.
%However, \cham 4) to which our work owes much can not predict
%Weinberg angle and Higgs mass because of the generation matrix  $K_{ij}$.
%Though above predictions are essentially in tree level, the possible
%quantum corrections are discussed by regarding the shift of $\delta$ from 1
% as the quantum effect.
\par
\noindent
\centerline{\bf   leptonic sector}
\par
Let us
reconstruct the fermionic sector of W-S theory using the formalism presented in
section \iii. We first consider the leptonic sector in the first generation.
Since leptons do not interact with gluons through the strong interaction,
we assign $\psi(x,0)=0$.
 In conformity with the chiral nature of leptons
we assign $SU(2)$ doublets
with left handed chirality on the second sheet $y=+$, and
$SU(2)$ singlets with right handed chirality on the third sheet $y=-$.
Consequently, in this context
$A^f_\mu(x,+)$ is given by
$A_\mu(x,+)$ in Eq.(36b) with $a\rightarrow a_{\mathop{}_{L}}$
and $A^f_\mu(x,-)$ by $A_\mu(x,-)$ in Eqs.(36c)
with $b\rightarrow a_{\mathop{}_{R}}$,
where $a_{\mathop{}_{L,R}}$ are related to hypercharges
of the leptons. In order that the $U(1)$ invariance is consistently realized
in Eqs.(34), $b-a=a_{\mathop{}_{R}}-a_{\mathop{}_{L}}$ is necessary.
In view of Eqs.(44) in which gauge fields and Higgs particle are scaled out,
we may set $a_{\mathop{}_{L,R}}=Y$ where $Y$ is the hypercharge of lepton.
With this in mind we make the following assignments for the leptons
in first generation.
$$
\psi(x,+)=l_{\mathop{}_{L}}(x)=\left(
          \matrix{
          \nu_{\mathop{}_{L}}(x)\cr
          e_{\mathop{}_{L}}(x)\cr
          }\right)\!,\quad a_{\mathop{}_{L}}=-1,
\eqno(49a)
$$
$$
\psi(x,-)=l_{\mathop{}_{R}}(x)=e_{\mathop{}_{R}}(x),\quad a_{\mathop{}_{R}}=-2,
\eqno(49b)
$$
where the left and right-handed spinors are defined by
$\psi_{\mathop{}_{L}}={{1-\gamma_5}\over 2}\psi$ and
$\psi_{\mathop{}_{R}}={{1+\gamma_5}\over 2}\psi$, respectively
with $\gamma_5=i\gamma^0\gamma^1\gamma^2\gamma^3$.
To be more precise the left-handed leptons form a doublet with
hypercharge $Y=a_{\mathop{}_{L}}=-1$
and
the right-handed
electron a singlet with
hypercharge $Y=a_{\mathop{}_{R}}=-2$ as usual.
After the scale transformations of gauge fields and Higgs field
as in Eq.(42), $a_{\mathop{}_{L,R}}$ is
nothing but the hypercharge $Y$.
Gauge invariance of Yukawa coupling (44)
is satisfied by $a_{\mathop{}_{L}}-a_{\mathop{}_{R}}=a-b=1$.
With these assignments, Dirac lagrangian for leptonic sector can be written as
$$
\eqalign{
        {\cal L}_{\mathop{}_{D}}(x)=&i\,[\,\,{\bar l}_{\mathop{}_{L}}(x)
            \, \gamma^\mu\,(\partial_\mu-{i\over 2}g\sum_i \tau^iA_\mu^i(x)
                       +{i\over 2}g'\tau^0B_\mu(x)\,)\,l_{\mathop{}_{L}}(x)\cr
                &\hskip 3.5cm +{\bar l}_{\mathop{}_{R}}(x)\,
        \gamma^\mu\,(\partial_\mu+ig' B_\mu(x)\,)\,l_{\mathop{}_{R}}(x)\,\,]\cr
&-g_{\mathop{}_{Y}}[\,\,{\bar l}_{\mathop{}_{L}}(x)\,H(x)\,l_{\mathop{}_{R}}(x)
    +{\bar l}_{\mathop{}_{R}}(x)\,H^\dagger(x)\,l_{\mathop{}_{L}}(x)\,\,]\,.\cr
}
 \eqno(50)
$$
The leptonic mass term is simply
$$\eqalign{
{\cal L}_{{\rm lepton}\;{\rm mass}}=&-g_{\mathop{}_{Y}}\,
   [\,({\bar\nu}_{\mathop{}_{L}}, {\bar e}_{\mathop{}_{L}})
     M(+)e_{\mathop{}_{R}}+{\bar e}_{\mathop{}_{R}}M(-)\left(
                              \matrix{
                              \nu_{\mathop{}_{L}}\cr
                              e_{\mathop{}_{L}}\cr
                              }\right)]\cr
&=-m_e({\bar e}_{\mathop{}_{L}}e_{\mathop{}_{R}}
      +{\bar e}_{\mathop{}_{R}}e_{\mathop{}_{L}})=-m_e\bar ee\,.\cr
}
\eqno(51)
$$
where $m_e=g_{\mathop{}_{Y}}\mu$.
Electron becomes massive whereas the neutrino remains massless as expected.
Inclusion of three generation is very easy for leptonic sector and so
we skip it.
\vfill\eject
%\par
%\noindent
\centerline{\bf quark sector}
\par
We consider the quark sector in three generations. In contrast to lepton,
quark interacts with gluon via the strong interaction and up-quark is massive.
In addition to these, quark mixing between three generations is taken into
 account. If we put $q_{\mathop{}_{F}}^{'}=(d^{'}, s^{'}, b^{'})$
and $q_{\mathop{}_{f}}=(d, s, b)$  quark mixing is expressed as
$$
          q_{\mathop{}_{F}}'=\sum_{f=1}^3 U_{\mathop{}_{Ff}}q_f, \eqno(52)
$$
where $U$ is Kobayashi-Maskawa mixing matrix.
In order for up-quark and down-quark to be massive, two kind of assignment are
necessary. The first assignment is to give the down-quark
$$
\eqalign{
     \psi(x,0)&=0, \cr
    \psi(x,+)&=a_1\,q_{\mathop{}_{L}}=a_1\left(
          \matrix{
          u_{\mathop{}_{L}}(x)\cr
          d'_{\mathop{}_{L}}(x)\cr
          }\right),\quad a_{\mathop{}_{L}}={1 \over 3},\cr
     \psi(x,-)&= d'_{\mathop{}_{R}}(x),
             \quad a_{\mathop{}_{R}}=-{2 \over 3},\cr}
\eqno(53)
$$
where we abbreviate subscripts about triplet representation of $SU(3)_c$ and
$a_1$ exists for normalization of quark fields in the final expression.
We have to carefully investigate the differential representation $A^f(x,y)$
in eq.(34). $\psi(x,+)$ is triplet for $SU(3)_c$, doublet for $SU(2)$
and of course singlet for $U(1)$
and transformed according to $g(x,0)\otimes g^f(x,+)\,$,
where  $g^f(x,+)$ is given as $e^{-ia_{\mathop{}_{L}}\alpha(x)}g(x)$.
$\psi(x,-)$ is triplet for
$SU(3)_c$ and singlet for $SU(2)\times U(1)$ and transformed according to
$g(x,0)\otimes g^f(x,-)\,$,
where $g^f(x,-)$ is $e^{-ia_{\mathop{}_{R}}\alpha(x)}$.
As a result $A^f(x,y)$ is determined to be
$$
\eqalign{
    A^f(x,+)&=(-{i\over 2}g_s\sum_{a=1}^8 \lambda^aG_\mu^a(x)
    -{i\over 2}g\sum_{i=1}^3 \tau^iA_\mu^i(x)
                 -{i\over 6}g'\tau^0 B_\mu(x))dx^\mu
                 +g_{\mathop{}_{H}}H(x)\chi
                  ,\cr
    A^f(x,-)&=\,(-{i\over 2}g_s\sum_{a=1}^8 \lambda^aG_\mu^a(x)
              +{i\over 3}g'B_\mu(x)\,)\,dx^\mu
               +g_{\mathop{}_{H}} H^\dagger(x)\chi\,. \cr
    }\eqno(54)
$$
The associated spinor one form is taken to be

$$
{\tilde {\cal D}}\psi(x,y)= \gamma_\mu \psi(x,y)dx^\mu
              +ic_d\psi(x,y)\chi.
\eqno(55)
$$
Adding to this assignment we make one more assignment
to give mass to up-quark.
$$
\eqalign{
     \psi'(x,0)&=0, \cr
    \psi'(x,+)&=a_2\left(
          \matrix{
          d_{\mathop{}_{R}}^c(x)\cr
         - {u'_{\mathop{}_{R}}}^c(x)\cr
          }\right),
               \quad a_{\mathop{}_{R}}={-1 \over 3},\cr
     \psi'(x,-)&=u'^c_{\mathop{}_{L}}(x),
     \quad a_{\mathop{}_{L}}(x)=-{4 \over 3},\cr}
\eqno(56)
$$
where superscripts $c$ on quark fields represent the charge conjugation and
$a_1^2+a_2^2=1$ is satisfied for normalization of quark field.
$u'$ denotes the up-quark mixed by  $U^\dagger$ matrix. It should be noted
that $\psi'(x,+)$ is the right handed spinor and $\psi'(x,-)$, left handed.
$\psi'(x,+)$ is ${\underline {3}}^\ast$  representation
for $SU(3)_c$, doublet for $SU(2)$
and of course singlet for $U(1)$, and then transforms according to
$g(x,0)^\ast\otimes g^f(x,+)\,$,
where  $g^f(x,+)$ is given as $e^{-ia_{\mathop{}_{R}}\alpha(x)}g(x)$.
$\psi'(x,-)$ is ${\underline {3}}^\ast$  representation for
$SU(3)_c$ and singlet for $SU(2)\times U(1)$, and transforms according to
 $g(x,0)^\ast\otimes g^f(x,-)$,
 where $g^f(x,-)$ is $e^{-ia_{\mathop{}_{L}}\alpha(x)}$.
For this assignment, $A^{f'}(x,y)$ are
$$
\eqalign{
    A^{f'}(x,+)&=({i\over 2}g_s\sum_{a=1}^{ 8} \lambda^{\ast a}G_\mu^a(x)
    -{i\over 2}g\,\sum_{i=1}^3 \tau^iA_\mu^i(x)
                 \! +\!{i\over 6}g'\tau^0 B_\mu(x)\,)dx^\mu
                 \!+\!g_{\mathop{}_{H}}H(x)\chi ,\cr
    A^{f'}(x,-)&=(\,{i\over 2}g_s\sum_{a=1}^{8} \lambda^{\ast a}G_\mu^a(x)
           +{2i\over 3}g'B_\mu(x)\,)dx^\mu
            +g_{\mathop{}_{H}}H^\dagger(x)\chi. \cr
    }\eqno(57)
$$
The associated one form spinor is taken to be
$$
{\tilde {\cal D}}\psi'(x,y)= \gamma_\mu \psi'(x,y)dx^\mu
              +ic_u\psi'(x,y)\chi.
\eqno(69)
$$
With these considerations, we can obtain the final expression as follows.
$$
        {\cal L}^q = {\cal L}^{q}_{kin.}+{\cal L}^{q}_{\mathop{}_{Yukawa}},
\eqno(58a)
$$
where
$$
\eqalign{
{\cal L}^{q}_{kin.}&=i{\bar{q}}_{\mathop{}_{L}}\,\gamma^\mu\,
               (\,\partial_\mu  -{i\over 2}g_s\sum_{a=1}^8 \lambda^a\,G_\mu^a
    -{i\over 2}g\sum_{i=1}^3 \tau^iA_\mu^i
                 -{i\over 6}\,g'\,\tau^0\, B_\mu\,)\,q_{\mathop{}_{L}}\cr
            &+i{\bar u'_{\mathop{}_{R}}}\,\gamma^\mu\,
            (\,\partial_\mu-{i\over 2}g_s\sum_{a=1}^8 \lambda^a\,G_\mu^a
               -i{2\over3}g'B_\mu)\,u'_{\mathop{}_{R}}\cr
         &+ i{\bar d'_{\mathop{}_{R}}}\,\gamma^\mu\,
         (\,\partial_\mu-{i\over 2}g_s\sum_{a=1}^8 \lambda^a\,G_\mu^a
         +i{1\over3}g'B_\mu\,)\,d'_{\mathop{}_{R}}\,,\cr}
\eqno(58b)
$$
and
$$
{\cal L}^{q}_{\mathop{}_{Yukawa}}
       =-g_{\mathop{}_{Y}}\,(\,{\bar q}_{\mathop{}_{L}}\,H\,d'_{\mathop{}_{R}}
              + {\bar d'}_{\mathop{}_{R}}\,H\,q_{\mathop{}_{L}}\,)
  -g'_{\mathop{}_{Y}}\,(\,{\bar q'}_{\mathop{}_{L}}\,
              {\tilde H}\,u'_{\mathop{}_{R}}
       + {\bar u'}_{\mathop{}_{R}}\,{\tilde H}\,q'_{\mathop{}_{L}}\,)\,,
\eqno(58c)
$$
with $g_{\mathop{}_{Y}}$ and $g'_{\mathop{}_{Y}}$ appropriately chosen
 Yukawa coupling constants.
  $q'_{\mathop{}_{L}}$ and ${\tilde H}$ in Eq.(58c) is given as
$$
           q'_{\mathop{}_{L}} =\left(\matrix{
                           u'_{\mathop{}_{L}} \cr
                           d_{\mathop{}_{L}} \cr}
                           \right), \hskip 1cm
            {\tilde H}=i\tau^2H^\ast.
$$
For the brevity, we set $q'_{\mathop{}_{L}}=q_{\mathop{}_{L}}$.
To be accurate, Eq.(58a) is the lagrangian for the first generation.
We have add the lagrangians over the three generations
to obtain the full lagrangian.
\par
We have shown that it is possible to reconstruct standard model with the
symmetry  $SU(3)_c\times SU(2)\times U(1)$  spontaneously broken to
$SU(3)_c\times U(1)_{em}$
based on Model I.
%The upper limit of Higgs mass can be predicted in Model I
%though in the tree level.
%On the other hand, \cham 4) can not predict
%anything even in tree level because of the  generation matrix $K_{ij}$.
%This is because our theoretical framework is more akin
%to the ordinary differential geometry.
%\par
$a,b$ in Eqs.(36b,c) play an fairly important role.
%in reconstructing the theory.
$a, b$ are the hypercharge of the particle
after the rescaling of fields in Eq.(42). However, $a-b=a_{\mathop{}_{L}}-
a_{\mathop{}_{R}}$ is important in the particle assignments to preserve
$U(1)$ symmetry. Thus, Weinberg angle in Eq.(56) is written in terms of two
parameters $r={b\over a}$ and $\delta$. As shown hereafter, Model
I\hskip-0.2mm I can determine $r$ due to the physical requirement.
%and predicts the interesting relation between Higgs mass, W boson mass and
%Weinberg angle.
%This relation 9) has already gotten in the slightly different formalism.
The fermion sector is easily incorporated in the formalism.
Thus, we skip it.
\par \noindent
\leftline{ \bf Model I\hskip -0.2mm I}
\par
Gauge and Higgs fields are assigned as follows, characteristic to satisfy
so called  the custodial symmetry. The notations are same as in Model I.
$$
\eqalign{
    A_\mu(x,0)&= -{i\over 2}\sum_{a=1}^8 \lambda^a G_\mu^a(x),\cr
    \Phi(x,0)&=0, \hskip 1cm M(0)=0 \cr
    g(x,0)&=g_s(x),\hskip 1cm g_s(x)\in SU(3)_c,\cr}
    \eqno(36a)
$$
$$\eqalign{
A_\mu(x,+)&=-{i \over 2}\sum_i\tau^i A_\mu^i(x)-{i\over 2}a\tau^0B_\mu(x),\cr
\Phi(x,+)&=\Phi(x)=
\left(
\matrix{
\hfill\phi_+(x)&-\phi_0^{*}(x)\hfill\cr
\hfill\phi_0(x)&\phi_-(x)\hfill\cr
}\right),\cr
M(+)&=
\left(
\matrix{
\hfill 0    & -\mu \hfill\cr
\hfill \mu  &0    \hfill\cr
}\right), \cr
g(x,+)&=e^{-ia\alpha(x)}g(x),\;e^{-ia\alpha(x)}\in U(1),\;
g(x)\in SU(2). \cr
}
\eqno(59a)
$$
and
$$\eqalign{
A_\mu(x,-)&=-{i \over 2}
\left(
\matrix{
\hfill bB_\mu(x)&\;0\;\;\hfill\cr
\hfill 0\quad &\;0\;\;\hfill\cr
}\right),\cr
\Phi(x,-)&=
\Phi^{\dag}(x,+),\cr
M(-)&=M^{\dag}(+),\quad \cr
g(x,-)&=\left(
                                \matrix{
                                \hfill e^{-ib\alpha(x)}&0\;\; \hfill\cr
                                \hfill 0\quad &1\hfill\cr
                                }\right).\cr
}
\eqno(59b)
$$
In this assignment, the auxiliary fields are
$$
  Y(x,\pm) = \sum_i a_i^\dagger(x,\pm)\mu^2\tau^0 a_i (x,\pm)
      =\mu^2\tau^0,
$$
because
$$
        M(+)M(-)=\left(
                  \matrix{
                     \hfill 0    & -\mu \hfill\cr
                     \hfill \mu  &0    \hfill\cr
                      }\right)\cdot
                      \left(
                  \matrix{
                     \hfill 0    & \mu \hfill\cr
                     \hfill -\mu  &0    \hfill\cr
                      }\right)
                      =\mu^2\tau^0 = M(-)M(+).
$$
Thus, both auxiliary fields $Y(x,\pm)$ are constant fields, which allows
both Higgs potential terms to remain in \ymh .
This model is very interesting to enable us to reconstruct standard model
in the gauge invariant way, also in Ref 6).
\par
In order to determine $r={b\over a}$, the gauge boson mass term is investigated
from Eqs.(18) and (27). It is given to be the trace of
$$
\eqalign{
A_\mu(x,+)M(+)-&M(+)A_\mu(x,-)\cr
&=-{i\over 2}\mu\left(
     \matrix{
     \hfill A_\mu^1-iA_\mu^2\quad &\;-aB_\mu-A_\mu^3\hfill\cr
     \hfill (a-b)B_\mu-A_\mu^3 &\;-(A_\mu^1+iA_\mu^2)\hfill\cr
     }\right),\cr}
$$
multiplied by the Hermite conjugate of this term.
This equation yields $a-b=-a \rightarrow b=2a\;(r=2)$ since photon is massless.
Putting $r=2 $ and using the same notations in Model I, we get \ymh as
$$\eqalign{
{\cal L}_{\mathop{}_{YMH}}=&
-{1\over 4g_s^2}\sum_a G_{\mu\nu}^a(x)G^{a\mu\nu}(x)\cr
&-{1\over 4g^2}
( \sum_i F^i_{\mu\nu}(x)\cdot F^i_{\mu\nu}(x)
+
(1+2\delta^2)a^2B_{\mu\nu}(x)B^{\mu\nu}(x))\cr
&+{2\alpha^2\over g^2}(1+\delta^2)\Bigl((D_\mu H(x))^{\dag}(D^\mu H(x)\Bigr)
-{2\alpha^4\over g^2}(1+\delta^2)(H^{\dag}(x)H(x)-\mu^2)^2.\cr
}
\eqno(60)
$$
By changing the scale of fields as $G_\mu(x) \rightarrow g_s G_\mu(x),\;$
$ A^i_\mu(x) \rightarrow g A^i_\mu(x),\;
B_\mu(x)\rightarrow {g\over a\sqrt{1+2\delta^2}}B_\mu(x)$ ,
and $H(x)\rightarrow{g\over \sqrt{2+2\delta^2}\alpha}H(x)$,
$$\eqalign{
{\cal L}_{\mathop{}_{YMH}}=&
-{1\over 4}\sum_a G_{\mu\nu}^a(x)\cdot G^{a\mu\nu}(x)\cr
&-{1\over 4}
(\sum_i F^i_{\mu\nu}(x)\cdot F^{i\mu\nu}(x)
+
B_{\mu\nu}(x)B^{\mu\nu}(x))\cr
&+(D_\mu H(x))^{\dag}(D^\mu H(x))
-\lambda(H^{\dag}(x)H(x)
-\mu'^2\,)^2,\cr
}
\eqno(61)
$$
with
 $g^{'}={g \over \sqrt{1+2\delta^2}}$,
$ \lambda={\eta^2g^{2} \over 2(1+\delta^2)}$ and
 $\;\mu'={\sqrt{2(1+\delta^2)}\alpha\mu \over g}$
 \hfill\break
{}From these equations we can predict in Model I\hskip -0.2mm I,
$$
\eqalign{
&\sin^2{\theta_{\mathop{}_{W}}}={1\over 4}\cdot {2 \over 1+\delta^2}, \cr
&m_{\mathop{}_{H}}=\sqrt{2}\eta m_{\mathop{}_{W}}
     \cdot \sqrt{2 \over {1+\delta^2}}.\cr} \eqno(62)
$$
The first equation in Eq.(62) lead to the inequality
$\sin^2{\theta_{\mathop{}_{W}}}\leq {1\over2}$.
Elimination of $\delta$ from Eq.(62) yields
$$
   m_{\mathop{}_{H}}=2\sqrt{2}\eta
m_{\mathop{}_{W}}\sin{\theta_{\mathop{}_{W}}}.
\eqno(63)
$$
If we put $\eta=1$, Eq.(63) is written as
$m_{\mathop{}_{H}}=2\sqrt{2} m_{\mathop{}_{W}}\sin{\theta_{\mathop{}_{W}}}$
in tree level, which leads to
$m_{\mathop{}_{H}}=109.1GeV$ when
$\sin^2\theta_{\mathop{}_{W}}=0.233$, and
$m_{\mathop{}_{W}}=79.9 GeV$ as the experimental values.
It is interesting to investigate
under what conditions $\eta=1$ is realized in non-commutative geometry.
%We expect that the quantum effects does not change this relation so much.
%If so, this relation predicts
\par
\section{
\hskip 4cm
{\bf \S 5. Conclusions and Discussions }}
We reconstruct standard model based on non-commutative geometry on  discrete
discrete
 space $M_4\times Z_3$. The inclusion of strong interaction into formalism
 is nicely realized in both boson and fermionic sectors.
For this purpose, we prepare three sheets $(Z_3)$. Our framework is more akin
 to the ordinary differential geometry than that of Connes and \cham.
Although Higgs mechanism has been a mystery in particle physics, it
is nicely explained owing to the theory of
differential geometry on the discrete space $M_4\times Z_3$.
The unified picture of the gauge and Higgs fields as a generalized connection
is realized. It should be stressed that the method to introduce the Dirac
lagrangian in section \S 3 is original, not in any text book of differential
geometry.\par
Our starting point in this paper is Eq.(1), where the unidentified factor
$a_i(x,n)$ appears. The matrix $M(x,y)$ appears through the generalized
differential derivative. The gauge field and Higgs field are written
in terms of $a_i(x,y)$ and $M(y)$ in Eq.(13). We again quote it.
$$
  \eqalign{
    A_\mu(x,y) &\= \sum_{i}a_{i}^\dagger(x,y)\partial_\mu a_{i}(x,y), \cr
    \Phi(x,y) &\= \sum_{i}a_{i}^\dagger(x,y)\,(-a_i(x,y)M(y)
            + M(y)a_i(x,-y)).\cr
}  \eqno(64)
$$
which makes us imagine that $a_i(x,y)$ would be the more fundamental object
and the gauge and Higgs bosons were composed of them. In addition to this,
$a_i(x,y)$ is restricted to be $\sum_i a_i^\dagger(x,y)a_i(x,y)=1$ which
seems to be the normalization condition for the field.
It should be remarked whether $a_i(x,y)$ becomes the physical object
experimentally observed in future. \par
 We can reconstruct
\ws with the same number of parameters as in the standard model.
Thus, quantization is performed as usual, contrary to the comment of
$\grave{\rm A}$lvarez et al 11). However, it is fun to imagine the following
thing.
The inner product of two-form on the discrete space $Z_3$ is taken to be
$<\chi\wedge\chi,\;\chi\wedge\chi>=\beta^4$, whereas $<\chi,\chi>=-\alpha^2$.
$\alpha=\beta$ is natural in the commutative geometry. It is very interesting
to investigate under what conditions $\beta$ is related to $\alpha$
in the non-commutative geometry on the discrete space. If such a condition
corresponds to a certain symmetry in the discrete space, it is conceivable
 to investigate the response against the quantum effect.

\par
%The justification of the present theory depends on the mass level
%of Higgs particle
%experimentally observed in future. Higgs mass is usually estimated
%to be in the range from $200\;GeV$ to $1\;TeV$.
% The present theory predicts Higgs mass to be
% about $110\;GeV$ in assuming the stability
% of Eq.(62) or the inequality in Model I against quantum effects.
\par
\vskip 1cm
\centerline{\bf Acknowledgement}
\par
The author would like to
express his sincere thanks to
Professor M.~Morita and J.~Iizuka
for useful suggestion on the noncommutative geometry
and invaluable discussions and
Professors H.~Kase
and
M.~Tanaka for informing some references and
useful discussions.
\vskip 1cm
\centerline{ \bf
Appendix{\hskip 1cm Higgs-Kibble model}
}
\par
Ref.6) could not treat Higgs-Kibble model because of
the gauge non-invariant term appearing in the Higgs potential.
In this appendix we formulate the $SU(2)$ Higgs-Kibble model
in the gauge invariant way presented in this paper.\par
We prepare also in this model two sheets($y=\pm)$ on which we prescribe
$a_i(x,+)$ to be complex $2\times2$ matrix-valued function
and $a_i(x,-)$ to be merely a real-valued function
%constant
%to satisfy
satisfying the relation $\sum_i a_i^\dagger(x,-)a_i(x,-)=1$.
With this prescription we immediately get $A_\mu(x,-)=0$.
We start with the field strength Eq.(16) and
 assign $A_\mu(x,y), \Phi(x,y)$ and $M(y)$ the following obvious notations.
\hfill\break
\leftline{$\underline{y=+}$}
$$\eqalign{
A_\mu(x,+)&=-{i\over 2}\sum_{k=1}^3\sigma^k A^k_\mu(x) = -{i \over 2}
\left(
\matrix{
A_\mu^3(x) &A_\mu^1(x)-iA_\mu^2(x) \cr
A_\mu^1(x)+iA_\mu^2(x) & -A_\mu^3(x) \cr
}\right),\cr
\Phi(x,+)&=
\left(
\matrix{
\phi_+(x)\cr
\phi_0(x)\cr
}\right),\cr
M(+)&=
\left(
\matrix{
0\cr
\mu\cr
}\right),
\;\; g(x,+)=g(x),\;g(x)\in SU(2).\cr
}
\eqno(A1a)
$$
\leftline{$\underline{y=-}$}
$$\eqalign{
A_\mu(x,-)&=0,\cr
\Phi(x,-)&=
(\phi_-(x),\;\; \phi_0^{*}(x))=\Phi^{\dag}(x,+),\;\phi_-(x)
=\phi_+^{*}(x),\cr
M(-)&=(0,\;\; \mu)=M^{\dag}(+), \quad g(x,-)=1.\cr
}
\eqno(A1b)
$$
Inserting these equations into Eq.(27) we get
$$\eqalign{
{\cal L}_{\mathop{}_{YMH}}=&-{1\over 4g^2}\sum_k
{F^k}_{\mu\nu}(x)\cdot {F}^{k\mu\nu}(x)
+{(1+\delta^2)\alpha^2\over g^2}(D_\mu\varphi(x))^{\dag}D^\mu\varphi(x) \cr
&-{\alpha^4\over g^2}{\rm tr}(\varphi(x)\varphi^{\dag}(x)-Y(x,+))^2
-{\delta^2\beta^4\over g^2}(\varphi^{\dag}(x)\varphi(x)-Y(x,-))^2.\cr
}
\eqno(A2)
$$
Here we have introduced the notation
$$\eqalign{
F_{\mu\nu}^i(x)&=\partial_\mu A_\nu^i(x)-\partial_\nu A_\mu^i(x)
+\epsilon_{ijk}A_\mu^j(x)A_\nu^k(x),\quad i,j,k=1,2,3,\cr
\varphi(x)&=
\left(
\matrix{
\varphi_+(x)\cr
\varphi_0(x)\cr
}\right),\quad \varphi_0(x)=\phi_0(x)+\mu,\quad
\varphi_+(x)=\varphi_-^{*}(x)=\phi_+(x),\cr
D_\mu \varphi(x)&=[\partial_\mu-{i \over 2}\sum_k{\sigma^k}\cdot
{A}^k_\mu(x)]\varphi(x),\;\cr
Y(x,y)&=\sum_{i}a_{i}^\dagger(x,y)M(y)M(-y)a_{i}(x,y),\cr
}
\eqno(A3)
$$
After eliminating the independent auxiliary field $Y(x,+)$ and
rescaling of fields; $A_\mu^i(x)\rightarrow gA_\mu^i(x)$ and $\varphi(x)
\rightarrow {g\over {\sqrt{1+\delta^2}}\alpha}\varphi(x)$,
we find that
$$\eqalign{
{\cal L}_{\mathop{}_{YMH}}=&-{1\over 4}
\sum_k{F^k}_{\mu\nu}(x)\cdot {F}^{k\mu\nu}(x)
+(D_\mu\varphi(x))^{\dag}(D^\mu\varphi(x))\cr
&-\lambda(\varphi^{\dag}(x)\varphi(x)
-\mu'^2)^2,\cr
}
\eqno(A4)
$$
where
$$
 \eqalign{
 F_{\mu\nu}^i(x)&=\partial_\mu A_\nu^i(x)-\partial_\nu A_\mu^i(x)
  +g\epsilon_{ijk}A_\mu^j(x)A_\nu^k(x),\quad i,j,k=1,2,3,\cr
       D_\mu \varphi(x)&=[\partial_\mu-{i \over 2}g\sum_k{\sigma^k}
       \cdot {A}^k_\mu(x)]\varphi(x),\cr}
\eqno(A5)
$$
with $\lambda={\delta^{2}\beta^4g^{2} \over \alpha^4(1+\delta^2)^2}$
and $\mu'={\sqrt{1+\delta^2}\alpha\mu \over g^2}$.
If we put $\alpha=\beta$, we can get the inequlity
%$m_{\mathop{}_{H}}={2\sqrt{2}\delta \over1+\delta^2}
$m_{\mathop{}_{H}} \leq \sqrt{2}m_{\mathop{}_{W}}$.
\vskip 1cm
%\vfill\eject

\centerline{{\bf References}}

\def\pl{Phys. Lett. }
\def\np{Nucl. Phys.}
\def\ptp{Progr. Theo. Phys.}
\def\prl{Phys. Rev. Lett.}

\item{1)} S.~Weinberg, {\prl} {\bf 19} (1967) 1264. \hfill\break
A.~Salam, {\it Weak and Electromagnetic Interactions, Elementary
Particle Theory},
ed. N.~Svartholm (John Wiley and Sons, Inc., New York, London,
Sydney, 1968), p.367.
\item{2)}
For the review of technicolor model,
E.~Farhi and L.~Susskind, Phys. Rep. {\bf 74} (1981) 277,\hfill\break
K.~Yamawaki, Proceeding of 1989 Workshop on Dynamical Symmetry
Breaking at Nagoya, p.11,\hfill\break
G.~Chapline and N.S.~Manton, \np ~{\bf B184} (1981) 391.
\item{3)}A.~Connes,
p.9 in {\it The Interface of Mathematics and Particle
Physics}, ed. D.~G.~Quillen, G.~B.~Segal, and Tsou.~S.~T.,
Clarendon Press, Oxford, 1990. See also,
Alain Connes and J. Lott,
Nucl. Phys. {\bf B}(Proc. Suppl.) {\bf 18B} (1990) 57.
\item{4)} A.~H.~Chamseddine, G.~Felder and J.~Fr\"olich,
\pl$\;$ {\bf B296} (1992) 109; \ \np$\;$ B395 (1993) 672.\hfill\break
See also, A.~H.~Chamseddine and J.~Fr\"olich,
^^ ^^  SO(10) Unification in Non-Commutative Geometry ",
ZU-TH-10/1993, ETH/ \ TH/93-12, 12 March 1992.
\item{5)} A.~Sitarz, \pl,\ {\bf B308}(1993) 311.                 \hfill\break
See also, Jagiellonian Univ. preprint,TPJU$-$7/92
^^ ^^  Non-commutative Geometry and Gauge Theory on Discrete Groups ".
\hfill\break
H-G.~Ding, H-Y.~Gou, J-M.~Li and K.~Wu,
preprint, ASITP-93-23, CCAST-93-5,
^^ ^^  Higgs as Gauge Fields on Discrete Groups and Standard Models
for Electroweak and Electroweak-Strong Interactions ".
\hfill\break
 This Beijing group paper
 also reconstructed standard model by use of Sitarz' formalism
 developed in the above preprint.
%%%%%%%%%%%%%%%%%%%%%%%%%%%%%%%%%%%%%%%%%%%%%%%%%%%%%%%%%%%%%%%%%%%%%%%%%
 The authors are thankful to Professor T.~Saito for informing
 this reference to us.
%%%%%%%%%%%%%%%%%%%%%%%%%%%%%%%%%%%%%%%%%%%%%%%%%%%%%%%%%%%%%%%%%%%%%%%%%
\item{6)} K.~Morita and Y.~Okumura, Nagoya Univ. preprint, DPNU-93-25,
\hfill\break
^^ ^^  Weinberg-Salam Theory in Non-Commutative Geometry ".
%\hfill\break
%submitted to \ptp.
\item{7)} K.~Morita and Y.~Okumura, Nagoya Univ. preprint, DPNU-93-38,
\hfill\break
^^ ^^   Reconstruction of standard model in non-commutative geometry on
$M_4\times Z_2$ ".
%\hfill\break
%submitted to \pr.
\item{8)} Y.~Okumura,
\hfill\break
$B!!(J^^ ^^  Gauge theory and Higgs mechanism based on differential geometry
  on discrete  space $M_4\times Z_{\mathop{}_{N}}\;$ ".
\item{9)} K.~Morita, \ptp $\;$ {\bf 90}(1993) 219.
\item{10)} K.~Morita and Y.~Okumura,
\hfill\break
^^ ^^  Weinberg-Salam Theory based on graded algebra in Non-Commutative
Geometry ".
\item{11)} E.~\'Alvarez, J.~M.~Gracia-Bond\'ia and
C.~P.~Martin, \pl$\;$ {\bf B306}(1993) 55.

\bye